\documentclass[usenatbib]{mnras}

\usepackage{amsmath}
\usepackage{graphicx}
\usepackage{amsfonts}
\usepackage{amssymb}
\usepackage{color}
\usepackage{bm}
\usepackage[T1]{fontenc} 
\usepackage{multirow}
\usepackage{float}
\usepackage{longtable}
\usepackage{amsmath,amssymb,amsfonts,mathrsfs,latexsym,graphicx}

\newcommand{\sigmares}{\sigma_{\rm res}}
\newcommand{\normal}{\mathcal{N}}
\newcommand{\zhat}{\hat{z}}
\newcommand{\mB}[1]{{m}^\star_{B#1}}
\newcommand{\Cparams}{\mathscr{C}} 
\newcommand{\BAHAMAS}{{\em BAHAMAS}}
\newcommand{\dr}{d_r}
\newcommand{\dc}{d_\text{cut}}
\newcommand{\SNIas}{SNIae}
\newcommand{\Mnote}{M_0^\epsilon}

\title[Host Galaxy Projected Distances]{Projected distances to host galaxy reduce SNIa dispersion }

\author[Hill et al]{Ryley Hill$^{1,2}$,
H.~Shariff$^{1,3}$,
R.~Trotta$^{1,3,4}$\thanks{E-mail: r.trotta@imperial.ac.uk},
S.~Ali-Khan$^{1}$,
X.~Jiao$^{3,5}$,
Y.~Liu$^{1}$,
S.-K.~Moon$^{1}$,
\newauthor 
W.~Parker$^{1,6}$,
M.~Paulus$^{1,7}$,
D.A.~van Dyk$^{3,4,5}$,
L.B.~Lucy$^{1}$
\\
$^{1}$Astrophysics Group, Physics Department, Imperial College London, Prince Consort Rd, London SW7 2AZ\\
$^{2}$Department of Physics and Astronomy, the University of British Columbia, 6244 Agricultural Road, Vancouver, BC, V6T 1Z1 \\
$^{3}$Imperial Centre for Inference and Cosmology, Astrophysics Group, Blackett Laboratory, Prince Consort Rd, London SW7 2AZ \\
$^{4}$Data Science Institute, William Penney Laboratory, Imperial College London, London SW7 2AZ\\ 
$^{5}$Statistics Section, Mathematics Department,  Huxley Building, South Kensington Campus, Imperial College London,  London SW7 2AZ\\
$^{6}$Bay House school \& Sixth Form, Gomer lane, Gosport, Hampshire,  PO12 2QP \\ 
$^{7}$Department of Physics and Astronomy, University of Glasgow, Glasgow G12 8QQ 
}

\date{Accepted XXX. Received YYY; in original form ZZZ}

\pubyear{2015}

\begin{document}
\label{firstpage}
\pagerange{\pageref{firstpage}--\pageref{lastpage}}
	\maketitle
\begin{abstract}
We use multi-band imagery data from the Sloan Digital Sky Survey (SDSS) to measure projected distances of 302 supernova type Ia (SNIa) from the centre of their host galaxies, normalized to the galaxy's brightness scale length, with a Bayesian approach. We test the hypothesis that \SNIas\ further away from the centre of their host galaxy are less subject to dust contamination (as the dust column density in their environment is smaller) and/or come from a more homogeneous environment. Using the Mann-Whitney U test, we find a statistically significant difference in the observed colour correction distribution between \SNIas\ that are near and those that are far from the centre of their host. The local $p$-value is $3\times 10^{-3}$, which is significant at the 5 per cent level after look-elsewhere effect correction. We estimate the residual scatter of the two subgroups to be $0.073 \pm 0.018$ for the far \SNIas, compared to $0.114 \pm 0.009$ for the near \SNIas\ -- an improvement of 30 per cent, albeit with a low statistical significance of $2\sigma$. This confirms the importance of host galaxy properties in correctly interpreting SNIa observations for cosmological inference.
\end{abstract}


\section{Introduction}

The Nobel Prize for Physics 2011 was awarded for the discovery that the universe is accelerating. \cite{Riess:1998cb} and \cite{Perlmutter:1998np} used \SNIas\ as standardizable candles to infer the existence of an additional component in the energy density of the universe, now called dark energy. The number of SNIa observations since that seminal discovery has grown rapidly: we now have hundreds of spectroscopically confirmed \SNIas\ ~\cite[e.g.,][]{Astier:2005qq, WoodVassey2007,Amanullah2010Spectra,Kowalski2008Improved, Kessler2009Firstyear,Freedman:2009vv,Contreras2009Carnegie,Balland2009ESOVLT,Bailey2008Initial,Hicken2009CfA3,2012ApJ...746...85S,Rest:2013mwz,betoule2014}, which have been used to measure the distance modulus to $z\sim 1.9$~\citep{Jones2013}. This low-redshift probe of the expansion history of the Universe, coupled with the high-redshift measurements of the Cosmic Microwave Background anisotropies, has been a vital tool for determining the equation of state of dark energy, and to put constraints on modified gravity models.

\SNIas\ are a subclass of supernovae defined by the absence of H lines in their spectra and the presence of Si lines. The generally accepted understanding of a SNIa explosion is that of a CO white dwarf accreting material from a companion star. The gravitational pressure ignites a runaway thermonuclear reaction that leads to the catastrophic unbinding of the white dwarf. While this scenario is generally agreed upon in literature, the details of their formation, including the nature of the progenitor, as well as the exact explosion mechanism remain unclear. The diversity of \SNIas\ appears however to suggest that \SNIas\ are produced by more than one progenitor channel: no single channel (i.e, single degenerate scenario, where a CO white dwarf accretes mass from a non-degenerate companion start; or double degenerate scenario, the merging of two white dwarfes) can account for all of the available observations (see e.g.~\citealt{Maeda:2016uhj} for a recent review).  

Given the emerging support for the idea that the diversity in \SNIas\ observations can only be explained by postulating multiple sub-classes (perhaps even within the `Branch-normal' SNIa type of~\citealt{1993AJ....106.2383B}),  it becomes even more important to clarify any links between the \SNIas\ variability and their galactic environment.  Much work has thus been expended in studying the influence of the SNIa galactic environment onto its observable properties -- in particular, possible residual dependencies of the brightness and/or colour of the SNIa after the empirical standardization corrections have been applied.   
%
%
Empirical corrections are applied by linearly adjusting the SNIa's $B$-band peak magnitude for stretch~\citep{Phillips:1993ng,1999AJ....118.1766P} (slow declining \SNIas\ are brighter) and colour excess~\citep{Riess:1996pa,JhaRiess2007} (bluer \SNIas\ are brighter). After such corrections, the residual scatter around the Hubble diagram is reduced to $\sim 0.1$ mag, which is what enables the use of \SNIas\ as cosmological probes. 

Since the standardization procedure is empirical (although partially justified by theoretical models, e.g.~\citealt{Kasen:2006is}), much effort has gone into trying to establish whether the residual scatter can be further reduced by including other observable proxies of the SNIa's environment and/or progenitor channel. These include host galaxy mass (e.g. \citealt{lampeitl2010,Sullivan:2010mg}), star formation rate, metallicity, stellar population age, spectral lines width, host morphology and location within the host. For a recent review of environmental correlations, see \cite{Anderson:2015kfa}.

There is general agreement that \SNIas\ in more massive galaxies are (post corrections) brighter~\citep{Kelly:2009iy,Sullivan:2006ah,Sullivan:2010mg,2016MNRAS.457.3470C,Shariff2016}, although estimates of the difference range from $0.055\pm 0.022$ mag  \citep{Shariff2016} to 0.11 mag~\citep{Sullivan:2010mg}. This effect could be a reflection of the hosts' metallicity~\citep{Gallagher:2008zi}, given the well-known mass-metallicity correlation in early-type galaxies. Several studies have investigated the influence of galaxy morphology and/or star formation rate (SFR), reporting that \SNIas\ with a smaller stretch parameter (i.e., rapid decliners) occur more often in lenticular/elliptical galaxies~\citep{Henne:2016mkt} as well as in passive galaxies~\citep{lampeitl2010}. \cite{lampeitl2010} found evidence that galaxies with lower star formation rates produce, on average, dimmer \SNIas. \cite{2011ApJ...743..172D} analysed host-galaxy spectra to obtain metallicity and star-formation rates from a subset of the {SDSS-II} \SNIas. {They found that \SNIas\ in high metallicity host galaxies are $\sim 0.1$ mag brighter post-correction, and a $>3 \sigma$ correlation between specific star formation rate and Hubble residuals. More recently, a study by \cite{2016MNRAS.462.1281M} investigated gas-phase oxygen abundance in 28 host galaxies in the local Universe, and concluded that high metallicity galaxies host dimmer \SNIas, in contrast to \cite{lampeitl2010}. However, given that \cite{2016MNRAS.462.1281M} did not apply colour corrections, the two results are not directly comparable.}

No significant correlation has been found between the \SNIas' colour and the hosts' morphology or SFR. For example,  \cite{2013ApJ...763...88C}  used a sample of 581 photometric \SNIas\ and applied a series of host galaxy corrections to the Hubble distance including metallicity, mass, star formation rate, specific star formation rate and age of galaxies. They found a $>5\sigma$ significant reduction in the Hubble residuals when host galaxy mass was added as an additional covariate. However, the improvement was less significant for the other covariates. More recently, much work has focused on evaluating the residual effect of the local SFR at the site of the explosion, with somewhat contradictory results~\citep{Anderson:2014aaa,Rigault:2013gux,Kelly:2014hla,Jones:2015uaa}.



Establishing the dependence (if any) of the \SNIas\ inferred luminosity on environmental effects is important both in terms of potentially improving \SNIas\ as standard candles and in order to reduce any remaining systematic effects.  For example, \cite{Kelly:2014hla} demonstrated that \SNIas\ in regions of high UV flux have a smaller residual dispersion. Furthermore, reducing the remaining intrinsic dispersion below 0.1 mag would enable measuring spatial weak lensing correlations between \SNIas~\citep{2014ApJ...780...24S}, a new probe of cosmological parameters~\citep{Scovacricchi:2016ylt}.

With this in mind, the goal of this paper is to revisit the question of the influence of the projected radial position of the SNIa within its host, but using a larger sample of \SNIas\ and a more sophisticated statistical analysis than was previously available.   

In this work we extend and improve previous (null) results by~\cite{ivanov2000,Yasuda:2009sc,galbany2012}, who analyzed \SNIas\ samples containing between 62 to 195 objects. 
\cite{ivanov2000} used a sample of 62 \SNIas\ and found no evidence of correlation between stretch or colour corrections and deprojected galactocentric distances from the hosts. They also separated the SNIa sample according to galactic morphology, again finding no effect.
\cite{Yasuda:2009sc} looked at the first year SDSS-II SN sample (137 \SNIas\ at $0.05 \leq z \leq 0.3$) and considered possible correlations of colour $c = E(B-V)$ with galactocentric distance, without finding any significant correlation. More recently, \cite{galbany2012} performed a similar analysis on a larger sample containing 195 \SNIas\ and again did not find an effect.

One of the motivations for the above studies was to investigate the effect (if any) of the radial metallicity gradient in the galaxy on the SNIa's properties. By selecting a sub-group at high projected galactocentric radius, one might hope to select a more homogeneous metallicity environment, and hence a more homogeneous sub-class of SNIa progenitors. Furthermore,  the effect of dust and extinction within the host also changes with galactocentric distance. The confounding effect of dust along the line of sight is a well-known source of uncertainty (and potentially systematic errors) in the empirical correction procedure.  Indeed there is evidence that the conventional linear colour correction is inadequate, in that it fails to distinguish between intrinsic colour variations in the \SNIas\ and host galaxy dust effects~\citep{Mandel:2016rks}. This is one for the reasons of the current effort to obtain light curve in the rest-frame Near Infrared -- a wavelength range much less affected by dust~\citep{2009ApJ...704..629M}.

However, given the high observational cost of obtaining NIR lightcurves for a sample at cosmological distances, it would be interesting to determine whether the location of the SNIa explosion within the host can help select \SNIas\ that are less affected by dust. While it is important to keep in mind that projection effects and host-galaxy selection biases will always be an issue, this could still lead to higher precision (by using \SNIas\ with lower residual scatter) and accuracy (by mitigating a potential source of systematic error).

Higher extinction is expected in regions with a higher star formation rate, and particularly in more central regions of star-forming galaxies. Hence, \SNIas\ at small galactocentric radii are expected to be redder, an effect that could be due to the local ISM but also to the progenitor's rings of dust, particularly for active, late-type galaxies. Indeed,~\cite{Anderson:2014aaa} showed that redder SNIa events are found more centrally in a sample of star-forming galaxies. They used the equivalent width of the unresolved sodium doublet (NaD) to quantify the amount of extinction along the line of sight, finding that \SNIas\ with NaD detections (indicating high absorption) are much more likely to be centrally located within their hosts. Largely removing the effect of reddening due to the ISM might help in identifying intrinsic colour variations in the \SNIas\ and/or the reddening due to Circumstellar Material (CSM). 
Therefore, our aim is to investigate whether segregating the SNIa sample into sub-groups according to their projected radial distance from the host can help in selecting a sub-group (at large galactocentric radii) that is less affected by dust and more homogeneous in its post-correction magnitude. Indeed, \cite{1996MsT..........3L} predicted that low reddened \SNIas\ would have very little interstellar gas and dust. This could be either due to the distance from host or the galaxy type \citep{2011Sci...333..856S}. 
As forcibly argued by~\cite{Mandel:2016rks}, the empirical corrections to the observed magnitude currently implemented in the popular {\em SALT2} fitter might introduce biases in the measured distances, for they bundle together intrinsic colour variations with extrinsic (i.e., due to host dust) variations, especially so at both extremes of the colour range. If it were possible to identify a sub-group of \SNIas\ that are less affected by extrinsic colour variations, this would be a useful tool to access their intrinsic colour variability and use such measurements as cross checks of systematics due to inappropriate statistical modeling of the type pointed out by~\cite{Mandel:2016rks}.

In this paper, we increase the sample size significantly with respect to previous work, to 302 \SNIas\ (a sub-set of the 368 SDSS \SNIas\ in JLA); we re-fit galaxy images from scratch (for consistency) and adopt a more sophisticated (and powerful) statistical approach for the estimation of the residual scatter after empirical corrections.

\vspace{20pt}

This paper is structured as follows. 
In Section \ref{sec:host} we discuss details of the modeling of the brightness profile of the host galaxies, how we estimate the \SNIas' galactocentric distance to their host, as well as how we split the \SNIas\ into two groups based on this quantity. In Section \ref{sec:BAHAMAS} we briefly summarize \BAHAMAS, the Bayesian parameter inference procedure we adopt to estimate the residual scatter in the two sub-groups of \SNIas.
We present our results in Section \ref{sec:results} and conclude in Section \ref{sec:conclusion}.

\section{Host Galaxy Modelling and Fitting}
\label{sec:host}

\subsection{Host Images Data}

We investigate the 368 \SNIas\ from the Sloan Digital Sky Survey (SDSS) II Supernova Survey~\citep{Sako2014} contained in the Joint Light-curve Analysis (JLA) SNIa compilation \citep{betoule2014}, {spanning the redshift range $0.04 \leq z \leq 0.4$ and with median redshift $z_\text{med} = 0.2$.} We require estimates of the SALT2~\citep{Guy:2007dv} parameters of each SNIa, namely the peak B-band magnitude, $m_B$, the lightcurve stretch correction, $x_1$, and colour correction, $c$. For these quantities we adopt the estimates obtained by~\cite{betoule2014}.  

\begin{figure}
\begin{center}
\includegraphics[width=\linewidth]{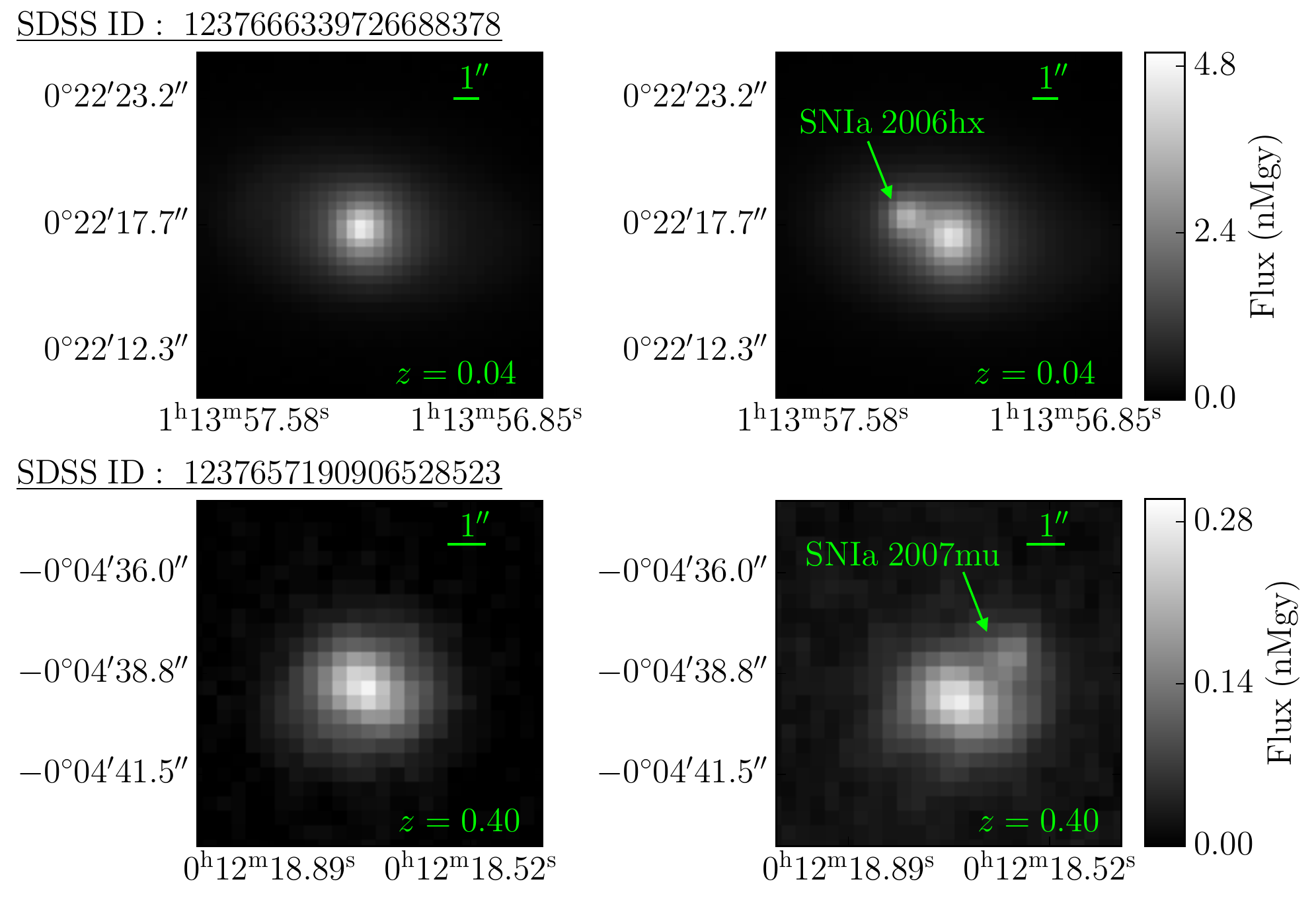}
\caption{Left: example galaxy images in the $r$-band, in units of nMgy (1\,nanomaggy$\,{=}\,3.631\,{\times}\,10^{-6}$\,Jy). The upper galaxy is a low redshift example while the lower galaxy is a high redshift example. Right: images of the corresponding \SNIas\ taken from the SDSS-II Supernova Survey. The \SNIas\ images were not used to fit galaxy intensity profiles, but only serve as illustrative examples.}
\label{fig:galaxy_stack}
\end{center}
\end{figure}

We obtain the necessary imaging to identify the position of each SNIa in its host galaxy from the SDSS Data Release 10 (DR10) \citep{ahn2014}. Images of $100 \times 100$ pixels were acquired, centred on the right ascension and declination of each host galaxy, as identified by the SDSS-II Supernova Survey. Given the pixel size of the SDSS, this resulted in square images of approximately 396 arcsec sides. Each galaxy was observed in five filters ($\textit{u}$$\textit{g}$$\textit{r}$$\textit{i}$$\textit{z}$) over the course of many different runs \citep{doi2010}. To obtain cutouts for fitting, we stacked the $r$-band images from all observations for a given galaxy with equal weights and normalised them to obtain a single co-added image for each galaxy, with arbitrary flux units, then further cropped to a size of $49 \times 49$ pixels around the reported SDSS host galaxy centre. An example of a galaxy image resulting form this procedure is shown in Fig.~\ref{fig:galaxy_stack}.

With increasing redshift of the galaxies, each of the bands is measuring bluer restframe wavelengths, and therefore the apparent size of the galaxy increases in a given band.  \cite{galbany2012} find (using $r$-band photometry for the SDSS galaxies) that the average spiral galaxy size increase by $\sim 30$ per cent up to $z\sim0.1$, and stabilizes after that. The same study did not find that this effect had an appreciable impact on their measurement of scaled SN distances from hosts. Furthermore, only 9 per cent of our hosts have redshifts less than 0.1, so it therefore appears safe to neglect this effect in the present study.

\subsection{Galaxy Images Fitting}

Our aim is to obtain approximate isophotes for the host galaxies and use these as a simple proxy for their dust column density. We are interested in determining the scale length of the galaxy, i.e., the projected distance to locus of elliptical isophote containing a specified fraction of the galaxy's flux (in the following, 50 per cent). The scale length then gives the characteristic scale by which to measure the projected distance (i.e., galactocentric distance) of the SNIa. 

Because obtaining an estimate of the scale length does not require sophisticated models for the light emission from the galaxy, we model each galaxy's radial intensity profile, $I(r)$, with a simple \citep{sersic1963} profile, as implemented in ProFit~\citep{Robotham:2016btg} Bayesian profile modelling code:
\begin{equation}
\label{eq:Ir}
I(r)=I_{e} \exp\left(-b_{n} \left[ \left( \frac{r}{r_{e}} \right) ^{1/n} -1 \right]\right),
\end{equation}
where 
\begin{align}
I_e & = \frac{10^{-0.4(m-m_0)}}{L_\text{tot}}, \\ 
L_\text{tot} & = \frac{2\pi A_\text{rat} r^2_e n \Gamma(2n)\exp(b_n)}{b_n^{2n}}.
\end{align}
\noindent
Here, $r_{e}$ is the scale length or effective radius (i.e., the radius containing half the total flux), $I_{e}$ is the intensity of the galaxy at $r_{e}$, $n$ is the S\'ersic index (which controls the drop-off in intensity from the centre, with $n=0.5$ corresponding to a Gaussian profile, $n=1$ being exponential, and $n=4$ being de Vaucouleurs), $m$ is the total flux expressed in magnitudes, $m_0$ is the magnitude zero-point, $A_\text{rat}$ is the ratio of the minor to major axis (with $A_\text{rat} = 1$ corresponding to a circle), $\Gamma$ is the standard gamma function, and $b_n$ is a derived parameter that ensures the correct integration of the flux at $r_e$.

The origin of the pixel coordinate system that we use for each galaxy image coincides by construction with the galaxy centre as reported in the SDSS-II Supernova Survey~\citep{Sako2014}. However, for internal consistency we refit the centre of the galaxy, parameterizing it by its pixel coordinates $(x_{0},y_{0})$. Furthermore, we allow for elliptical isophotes with minor to major axis ratio parameterized by $A_\text{rat}$ and major axis orientation angle, $\theta$ (increasing counter-clockwise from $0^\circ$ being vertical). With this, the radial distance $r$ in Eq.~\eqref{eq:Ir} of a pixel with coordinates $(x,y)$, is given by
\begin{equation}
\label{eq:r}
r(x,y) = \left( x_{p}^{2}+\frac{y_{p}^{2}}{A_\text{rat}^{2}} \right)^{1/2},
\end{equation}
where
\begin{align}
x_{p} & = (x-x_{0})\cos(\phi)+ (y-y_{0})\sin(\phi), \\ 
y_{p} & = -(x-x_{0})\sin(\phi)+ (y-y_{0})\cos(\phi), \\
\phi & = \theta + \pi/2.  
\end{align}
Our model of the intensity profile of each host galaxy
thus includes seven free parameters, namely 
\begin{equation}
\Theta = \left \{  x_{0}, y_{0}, m, \theta, A_\text{rat}, , r_{e}, n  \right \}.
\end{equation}
The model images were convolved with a Gaussian PSF of FWHM 1.3 arcsec, matched to the median SDSS PSF. We also fitted any additional point sources close to the host galaxy to ensure that they would not bias the reconstruction of the host galaxy's shape. 
 
Following the recommendations by the authors of ProFit~\citep{Robotham:2016btg}, we adopt a more robust Student-T distribution for the likelihood function, with heavier tails than a standard Normal distribution. At each observed pixel intensity, $D_{ij}$, the standard deviation is estimated as $\sigma_{ij} \equiv \sqrt{D_{ij}}$ (using the high-count Gaussian approximation to the underlying Poisson distribution for each pixel). Then the likelihood is, up to an irrelevant normalization constant:
\begin{align}
\label{eq:loglikelihood}
x_{ij} & = \frac{\left[(D_{ij} - I(r(x_{i},y_{j})|\Theta)\right]^{2}}{\sigma^2_{ij}}, \\
\ln \mathcal{L}(\Theta) & = -\ln \sum_{i,j}  \frac{\Gamma\left( \frac{\nu+1}{2}\right)}{\sqrt{\nu\pi}\Gamma\left(\frac{\nu}{2}\right)} \left(1+\frac{x_{ij}}{\nu} \right)^{-\frac{\nu+1}{2}}.
\end{align}
We used ProFit's Markov Chain Monte Carlo (MCMC) component-wise hit-and-run Metropolis (CHARM) sampler (from the \texttt{LaplacesDemon} package) to obtain the posterior distribution for the host galaxy intensity parameters vector, $\Theta$, using priors as given in Table~\ref{tab:priors}. We then base our final inference on 3,000 posterior samples, obtained from a thinned final chain after burn-in has been removed and convergence achieved. We report Maximum A Posteriori (MAP) values of parameters, while their uncertainty was obtained as the standard deviation of the marginal posterior. 

\begin{table}
\begin{center}
\begin{tabular}{l l l l l}
\hline
Parameter & Symbol & Prior \\ \hline
Galaxy centre [px] & $x_0$, $y_0$ & $\mathcal{N}\big([0,0],$\\
& & \quad \ \ $\text{diag}(0.5^2, 0.5^2)\big)$ \\ 
Apparent mag. ($r$-band) & $m$ & $\mathcal{U}[10, m_0]$ \\ 
Major axis orientation & $\theta$ & $\mathcal{U}[0, \pi]$\\
\quad angle [rad] & & \\
Major to minor axis ratio & $A_\text{rat}$ & $\mathcal{U}[0.1, 1]$\\
Scale length [px] & $r_e$ & $\mathcal{U}[0.5, 25]$\\
S\'ersic index & $n$ & $\mathcal{U}[0.5, 10]$\\
\hline
\end{tabular}
\caption{Priors adopted on galaxy images fitting parameters. The pixel coordinate system is chosen to be centred on the galaxy coordinates reported by the SDSS-II Supernova Survey. $\mathcal{N}(\mu, \Sigma)$ denotes a multivariate Gaussian with mean $\mu$ and covariance matrix $\Sigma$; $\mathcal{U}[a, b]$ denotes a uniform distribution in the range $a$ to $b$.
\label{tab:priors}}
\end{center}
\end{table}%

\begin{figure*}
\begin{center}
\includegraphics[width=\linewidth]{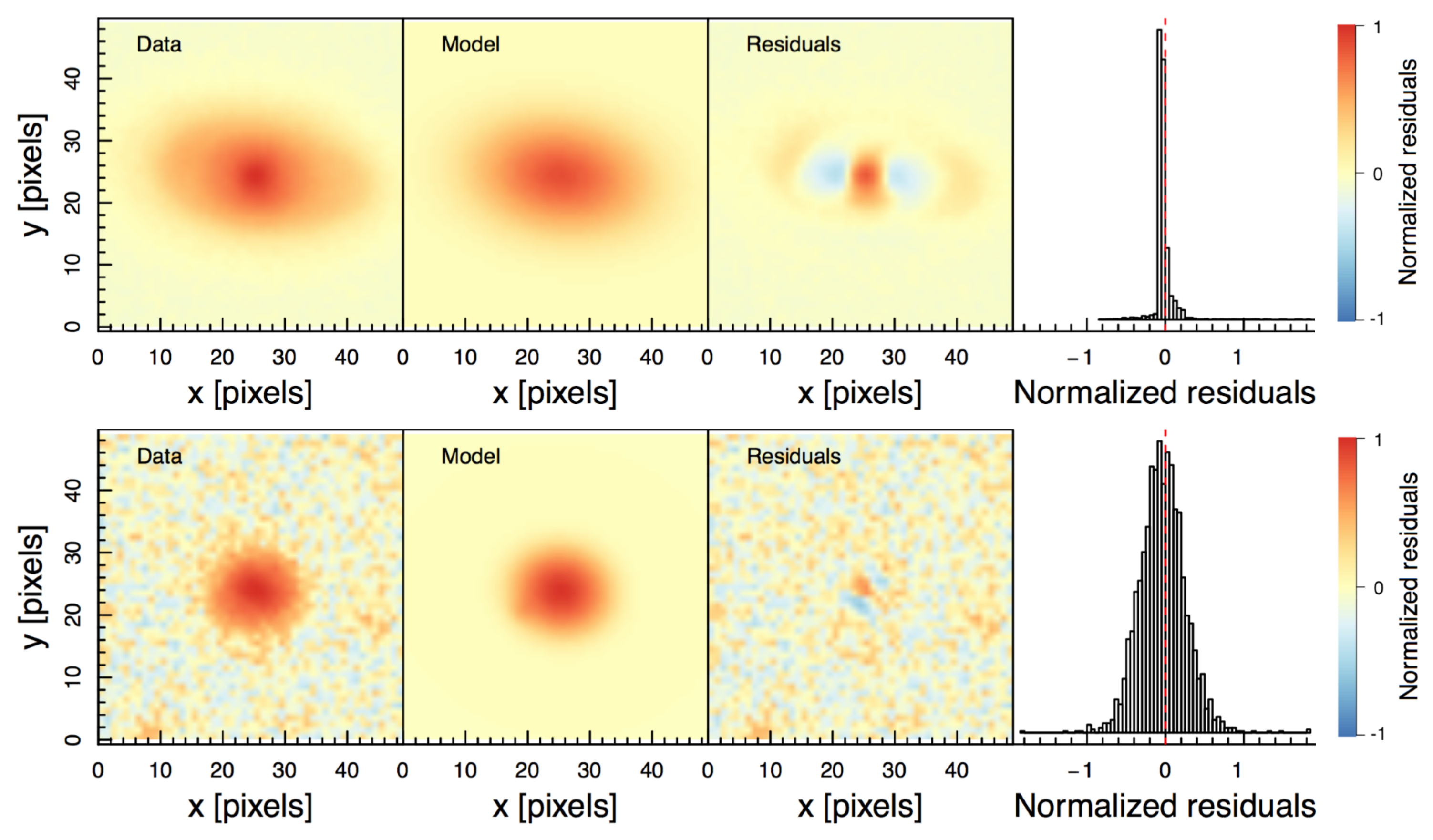}
\caption{ProFit host galaxy image modelling, showing the normalized SDSS $r$-band data (left panels), the ProFit MAP model (second panels) and the normalized residuals (right panels and histograms). The top row is for the host galaxy of SN2006hx ($z=0.04$, upper panel in Fig.~\ref{fig:galaxy_stack}) while the bottom row is for the host of SN2007mu ($z=0.40$, lower panel of Fig.~\ref{fig:galaxy_stack}).}
\label{fig:good_images}
\end{center}
\end{figure*}

We compared our MAP galaxy centre values with the host galaxy coordinates given by~\cite{Sako2014}: the median displacement is 0.27 pixels. A fit was deemed ``successful'' if (i) the iso-brightness contours followed the general shape of the galaxy, with $r_{e}$ being located somewhere in the vicinity of the outer edge of the galaxy; (ii) if $(x_{0},y_{0})$  was in the visual centre of the galaxy and compared well with the host galaxy coordinates in~\cite{Sako2014} and (iii) the residuals were close to 0 throughout the model image\footnote{Each fit is then visually inspected to ensure that the fitted luminosity profile is a good description of the image by checking the residuals plots.}  Two examples of a successful fits are shown in Fig.~\ref{fig:good_images}. While the top row shows some structure in the image residuals, it is clear that the model captures the overall shape of the galaxy's intensity sufficiently well. The bottom row shows an example of a higher redshift galaxy, with only little residual structure. 

\begin{figure}
\begin{center}
\includegraphics[width=\linewidth]{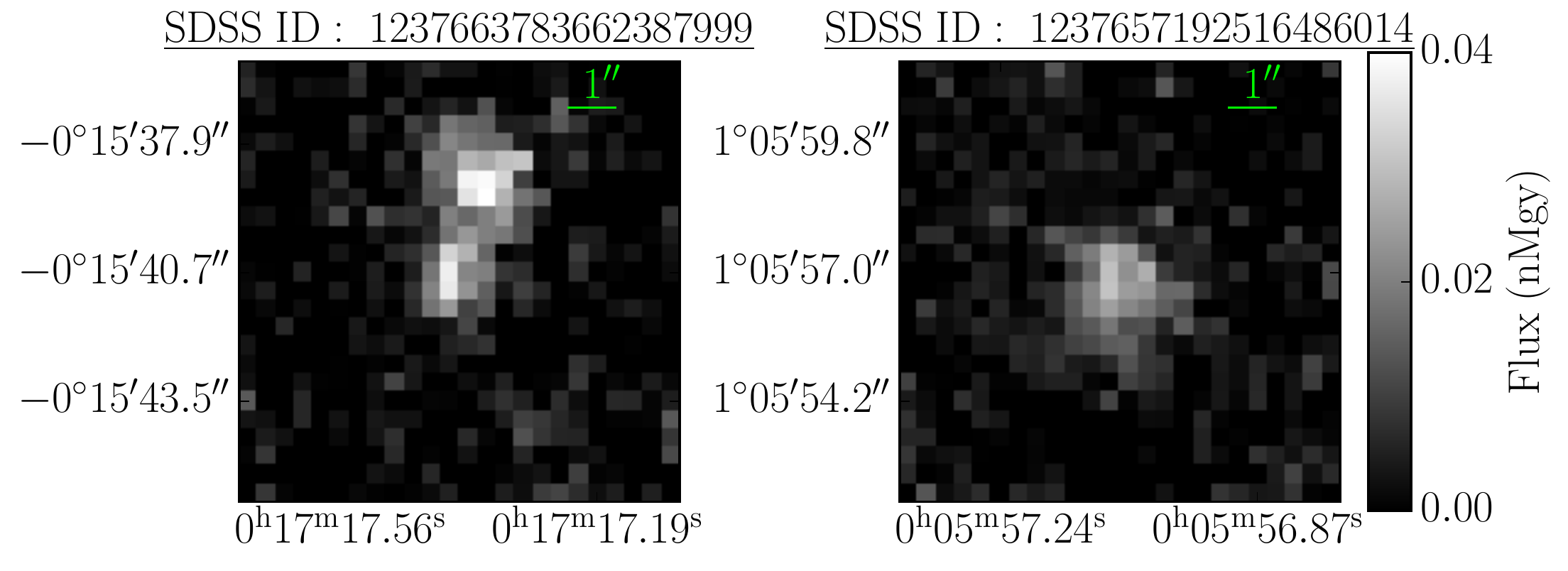}
\caption{Two examples of images that are of insufficient quality to be fit. \SNIas\ associated with galaxies that led to unsuccessful fits were not included in the analysis.}
\label{fig:poor_images}
\end{center}
\end{figure}

Some fits are unsuccessful because of poor image quality.  In some cases the galaxy is found in a region of significantly higher noise and in other cases the galaxy is too faint. Both of these effects lead to a host intensity within a few percent of the background. Examples of these cases are shown in Fig.~\ref{fig:poor_images}. \SNIas\ associated with a host galaxy that could not be fit are removed from the sample. 
Of the 368 host galaxy images, 66 cannot be successfully fit, leaving a sample of 302 \SNIas\ for our analysis. 

\begin{figure}
\begin{center}
\includegraphics[width=\linewidth]{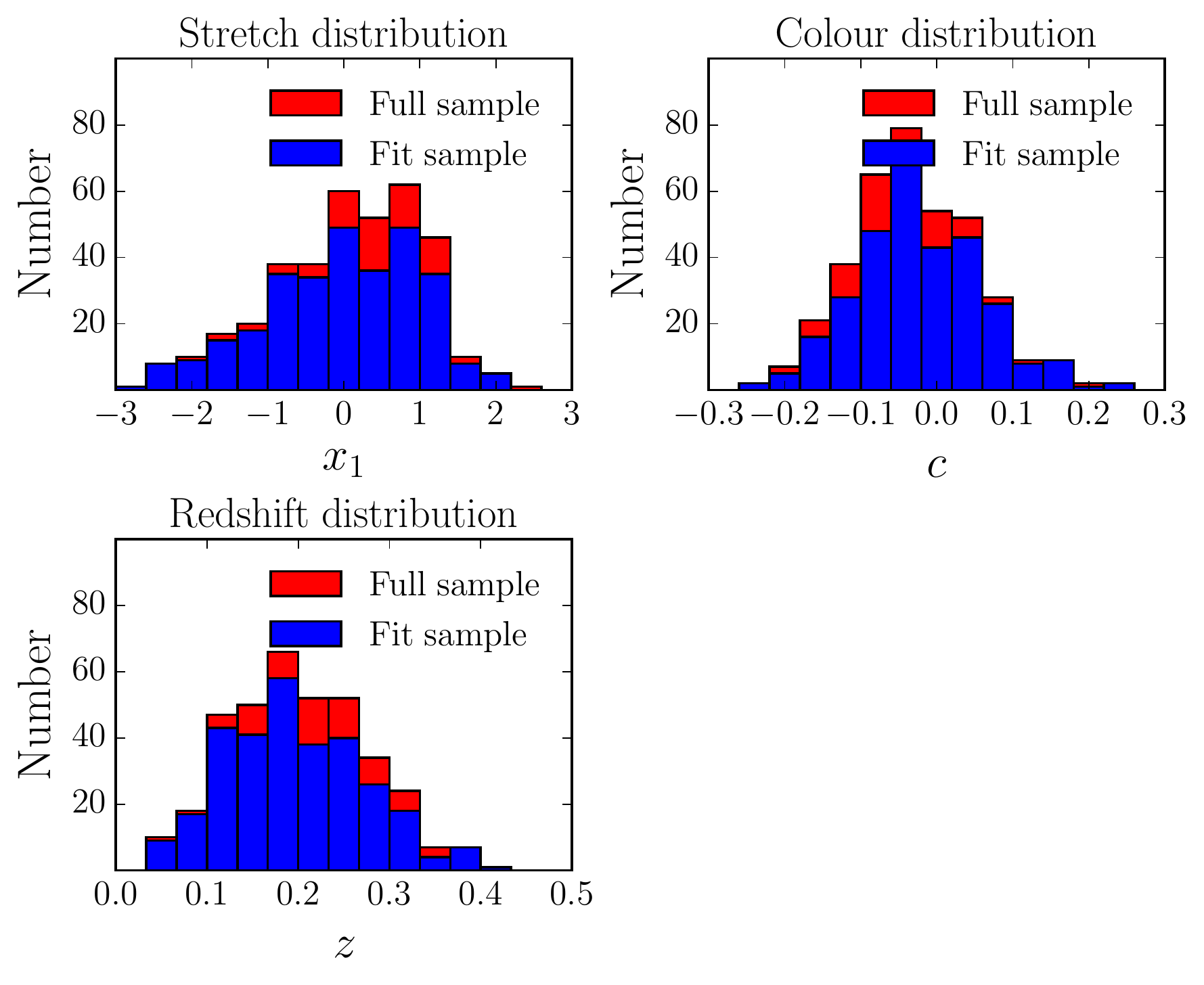}
\caption{The distributions of stretch, colour and redshift for the full host galaxy data set and after removing the SNIa with poor-quality host galaxy images which we deemed to be unreliable for fitting. The shapes of the distributions are similar, showing no evidence that our procedure has introduced bias in the distributions.}
\label{fig:cut_histogram}
\end{center}
\end{figure}

We checked our sample for any biases that may have been introduced by removing the poor fits; for example, it might be expected that higher redshift galaxies on average are fainter and thus tend to be more difficult to fit. In Fig.~\ref{fig:cut_histogram} we plot histograms of the stretch, colour and redshift (the 3 parameters which are relevant to cosmology and which we use in the following section) of the complete SDSS host galaxy sample, compared with the hosts that were successfully fit. We see that the shapes of the distributions remain virtually unaffected, which is compatible with the hypothesis that removing poor fits does not introduce bias to the distributions. 

\begin{figure}
\begin{center}
\includegraphics[width=\linewidth]{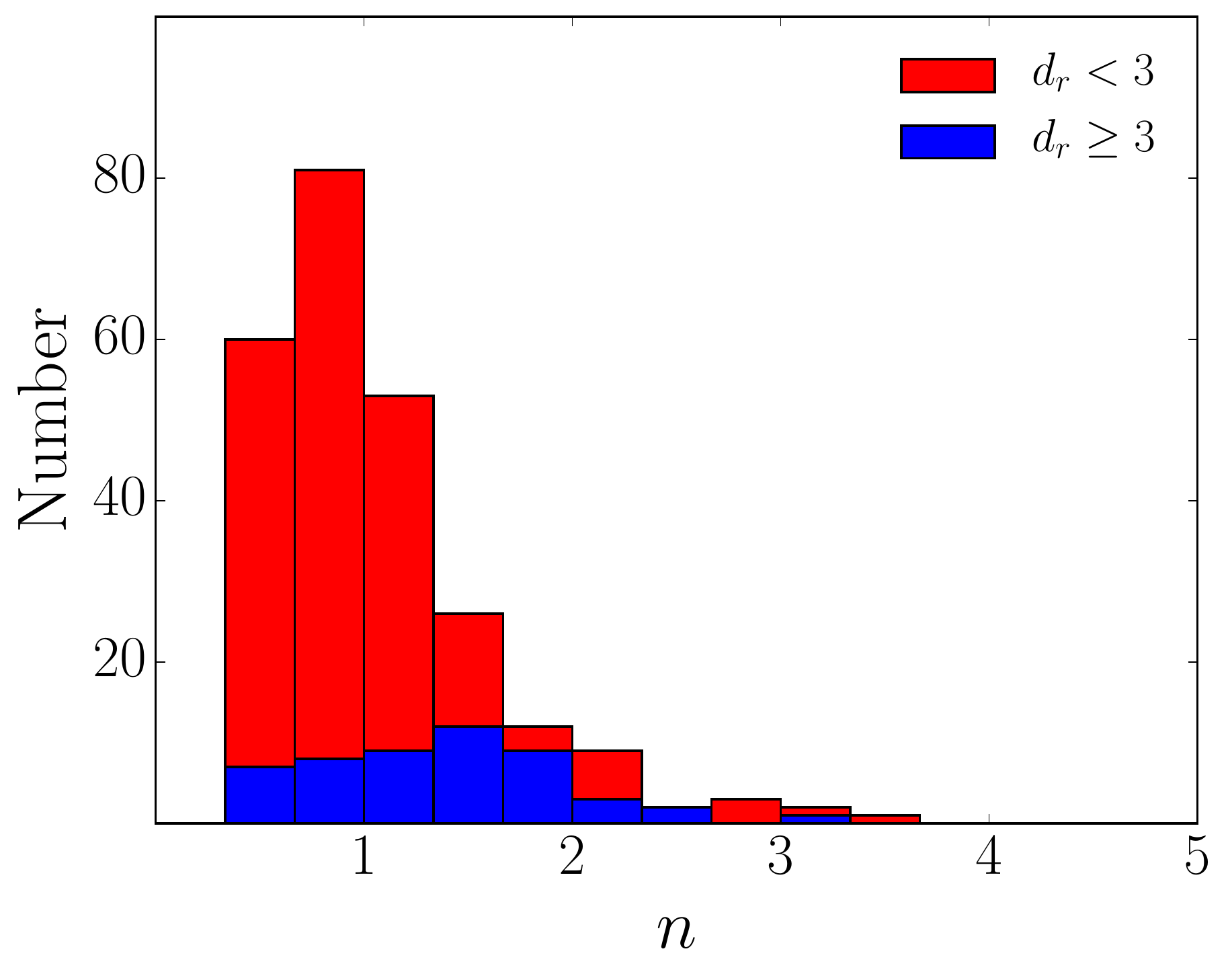}
\caption{{The distribution of host galaxies' reconstructed S\'{e}rsic parameters for the two groups of SNIae, $d_r<3$ and $d_r\geq 3$. The peak around $n=1$ corresponds to an exponential profile, known to be a good fit to disks in spiral galaxies; the absence of galaxies at $n \approx 4$ (an index associated with the profile of elliptical galaxies) can be attributed to cosmological dimming (i.e.~decreasing resolution and increasing noise) leading to more disk-like profiles \citep{paulino-afonso2017}.}}
\label{fig:ndistribution}
\end{center}
\end{figure}

{Historically, spiral galaxies have been identified with a S\'{e}rsic index $n=1$ (corresponding to a typical exponential disc galaxy) and elliptical galaxies with a S\'{e}rsic index $n=4$ \citep[a de Vaucouleur profile; see e.g.][]{devaucouleurs1959}. More recently, it has become standard to allow $n$ to vary freely, as there is significant spread around these integer values in modern surveys~\citep[e.g.,][]{ravindranath2004,trujillo2006,buitrago2008,vandokkum2010,vanderwel2012,vika2015}. Fig.~\ref{fig:ndistribution} shows the distribution of the inferred  S\'{e}rsic index $n$ for both sub-groups. 
The mode of the $d_r<3$ group is $n=1$, while the mode of the $d_r\geq 3$ group is $n=1.6$.  We observe that we have very few galaxies with $n>3$ and none with $n \approx 4$. This result is consistent with the findings of \citet{paulino-afonso2017}, who found that the reconstructed S\'{e}rsic indexes of 2507 local and 1118 high-$z$ star-forming galaxies are biased low by about 20\%, almost independently of redshift. They attributed this phenomenon to cosmological dimming, leading to decreased resolution and increased noise that produces a bias towards disk-like (i.e.~$n=1$) profiles. Nevertheless, we do not believe that this effect leads to a bias in our inferred values for the galactocentric distance, as explained in the next section.}

\subsection{Sub-groups of \SNIas\ based on galactocentric distance to host}
\label{sec:two_popn}

From our best-fit parameters we compute the projected distance of the SNIa from its host galaxy with Eq.~ \eqref{eq:r}:
\begin{equation}\label{eq:rsn}
r_{\rm SN} = r(x_{\rm SN}, y_{\rm SN}),
\end{equation}
\noindent
where $(x_{\rm SN},y_{\rm SN})$ are the coordinates of the SNIa in each host galaxy image determined from our fits (which generally compare well with the coordinates in~\citealt{Sako2014}), converted from standard RA/DEC coordinates with the {\em AstroPy} library for Python~\citep{astropy2013} which includes World Coordinate System (WCS) transformation functions \citep{calabretta2002}. This is the distance between an SNIa and its host galaxy if the SNIa were moved along an isophote contour to line up with the semi-major axis of the host. The galactocentric distance is obtained by dividing this distance by the scale length, $r_e$:
\begin{equation} \label{eq:impactparam}
d_{r} \equiv \frac{r_{\rm SN}}{r_{e}},
\end{equation}
\noindent
This definition of galactocentric distance is a good description of the SNIa to host galaxy distance as it takes into account the geometry of the light distribution of the host galaxy, which is assumed to be a proxy for the dust that is the main contributor to \SNIas\ extinction.

{The galactocentric distance defined in Eq.~\eqref{eq:impactparam} takes into account ellipticity and is normalized to the scale length $r_e$. By contrast, \citet{Sako2014} defined galactocentric distances using circular Petrosian half-light radii instead, making a direct comparison between our two methods difficult.}

\begin{figure}
\begin{center}
\includegraphics[width=\linewidth]{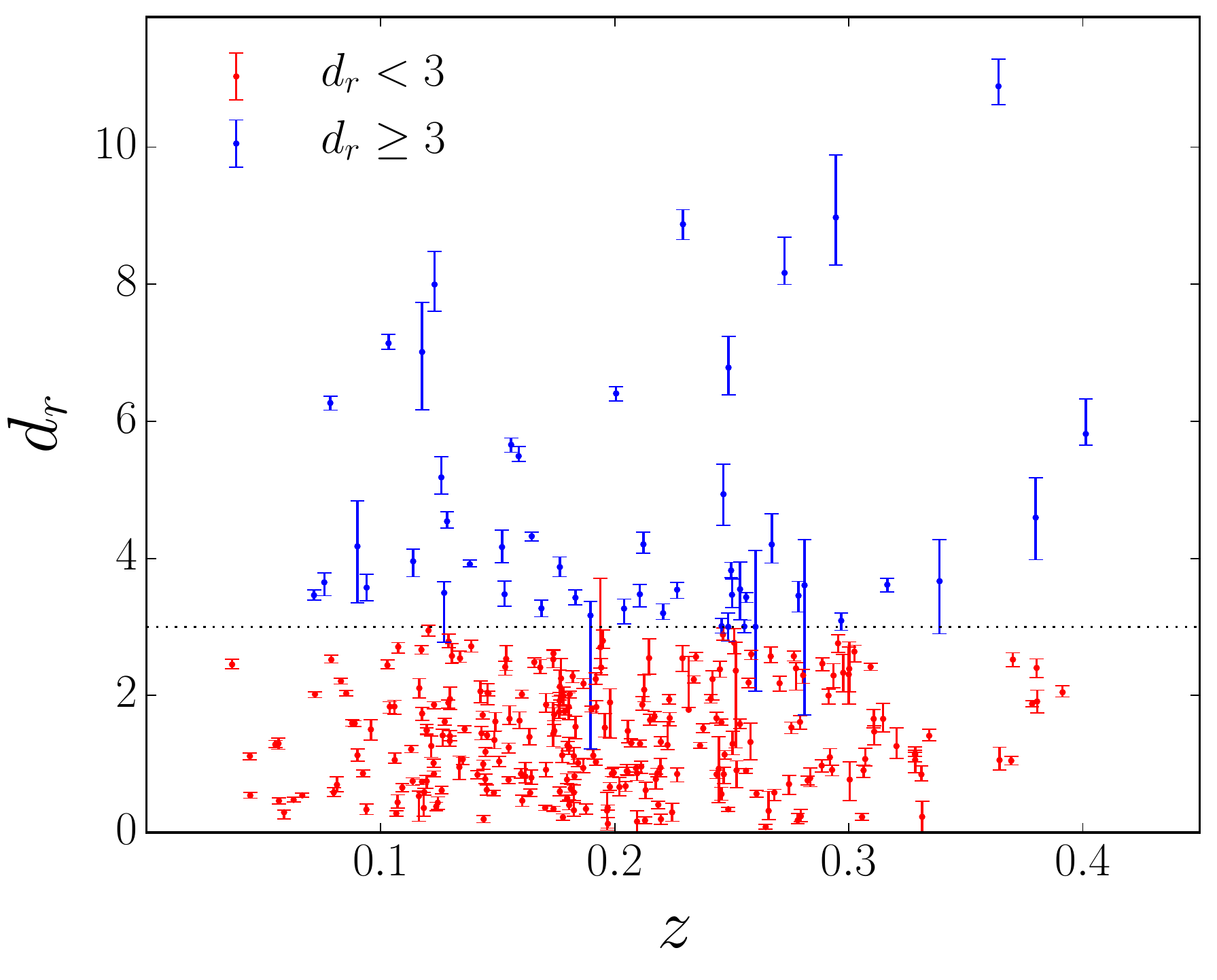}
\caption{Distribution of normalized SNIa-host galaxy distances calculated from our best-fit light profile model (Eq.~\ref{eq:impactparam}) as a function of redshift, where the uncertainties are obtained from calculating the 68 per cent credible range (Highest Posterior Density interval) of $d_{r}$ resulting from the MCMC fitting chain. The mean fractional error in our measurements is about 10 per cent. The horizontal dashed line at $d_{r}=3$ shows our chosen cut-off between high and low SNIa-host galaxy distances.}
\label{fig:dr_uncertainty}
\end{center}
\end{figure}

In Fig.~\ref{fig:dr_uncertainty} we show the galactocentric distances found for our sample of 302 galaxies as a function of redshift, where the uncertainties were estimated by calculating the 68\% credible intervals of the marginal posterior distributions resulting from ProFit's MCMC. We took into account the uncertainty of the SNIa positions during the fit by drawing their right ascensions and declinations from a Gaussian distribution with a width of 0.2 pixels, the typical pointing uncertainty of the SDSS telescope \citep{gunn2006}. We find that the mean fractional error in our measurements is about 10 per cent. While some uncertainties are quite large -- a result of poor image quality in several cases -- around the $\dc = 3$ value only relatively few galaxies have uncertainties that would put in question their classification in one of the two groups. Therefore we conclude that discarding such uncertainties is unlikely to lead to significant differences in the subsequent analysis.

{A potential difficulty would arise if our values of galactocentric distance $d_r$ were correlated with the S\'ersic index $n$. Since $d_r$ is inversely proportional to $r_e$, and $r_e$ is positively correlated with $n$~(see e.g.~\cite{trujillo2001}), a larger value of $d_r$ could correlate with a smaller value of $n$. If this were the case, our sub-groups could be a simple reflection of different galaxy morphology, rather than galactocentric distance. However, we find no significant correlation between $n$ and $d_r$. Fig.~\ref{fig:ndistribution}, showing the distribution of $n$ for both sub-groups of $d_r$, demonstrates that the mode of the $d_r\geq 3$ group is at a slightly higher value of $n$ than the mode for the $d_r < 3$ group. If $d_r$ and $n$ were anti-correlated, the larger $d_r$ sub-group distribution of S\'ersic indexes should peak at a lower value of $n$, not at a higher value like we observe. Therefore we conclude that our grouping based on $d_r$ is not a reflection of the host galaxies' morphology.}

\begin{figure}
\begin{center}
\includegraphics[width=\linewidth]{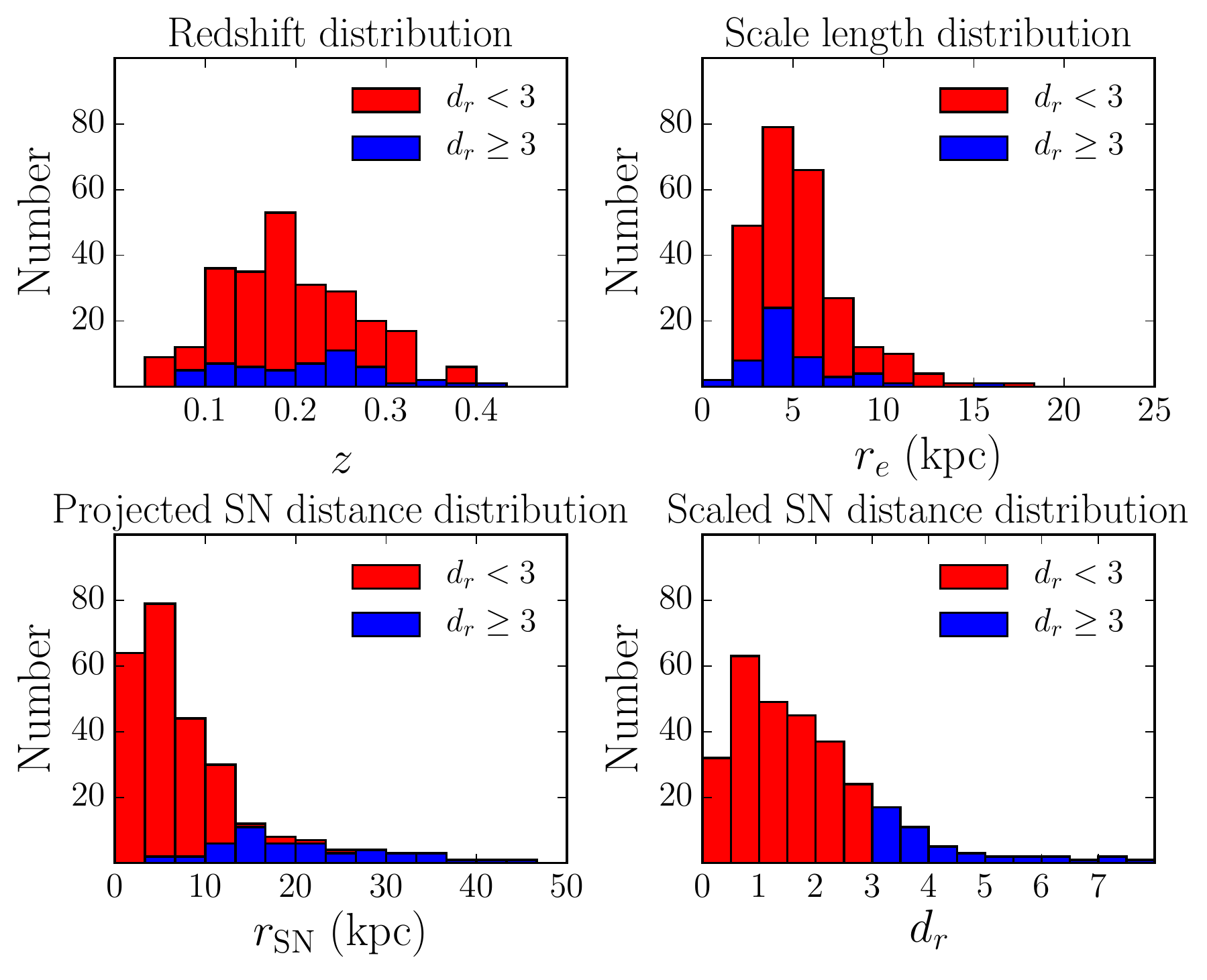}
\caption{The distribution of SNIa redshifts, host galaxy scale radii $r_{e}$, projected galactocentric distance of \SNIas, $r_\text{SN}$, and normalized galactocentric distances $d_{r}$. Here, $r_{e}$ is the length of the semi major axis of an ellipse centred on a host galaxy which extends out the elliptical contour enclosing half the total flux. $d_r$ is computed as the distance between a SNIa and the centre of its host galaxy if the SNIa were moved along a line of constant intensity to line up with the semi-major axis of the host, expressed in units of $r_{e}$.}
\label{fig:distr_dr}
\end{center}
\end{figure}

According to our hypothesis, \SNIas\ further from their galactic centre (i.e., with $\dr \gg 1$) should explode in a galactic environment with a smaller dust column density, and hence they should exhibit less dust absorption. Following~\cite{Anderson:2015kfa}, we also expect that \SNIas\ at large galactocentric distances ought to exhibit bluer values for the colour correction. We therefore expect the distribution of the colour correction parameter $c$, to differ for \SNIas\ with larger $\dr$ than for those with $\dr$ close to 0. In order to test this hypothesis, we split the sample of 302 \SNIas\ with $\dr$ values into two groups, one with $d_{r} < \dc$ and the other with $d_{r} \geq \dc$, where $\dc$ is a cut-off value. 

The choice of $\dc$ is important in establishing the significance (or otherwise) of the effect of $\dr$ on $c$. Given the distribution of $\dr$ in our sample, shown in Fig.~\ref{fig:distr_dr}, a larger value of $\dc$ is expected to give a cleaner sample (i.e., \SNIas\ less affected by dust) but at the price of a very small sample size. On the other hand, choosing a smaller cut-off value may lead to a substantial number of \SNIas\ with $\dr \geq \dc$ that are still affected by dust. We chose a value of $\dc = 3$ for our analysis, as it is reasonable to assume a transition in the amount of dust at a galactocentric distance more than twice the value of the scale radius of the galaxy. We emphasize that we are not choosing the value of $\dc$ to maximize the significance of the effect, but we still {\em a posteriori} check the significance of our results as a function of $\dc$ and correct our test statistics for multiple testing (i.e., there is no ``look elsewhere effect''), as described below.

\begin{figure}
\centering
	\includegraphics[width=\linewidth]{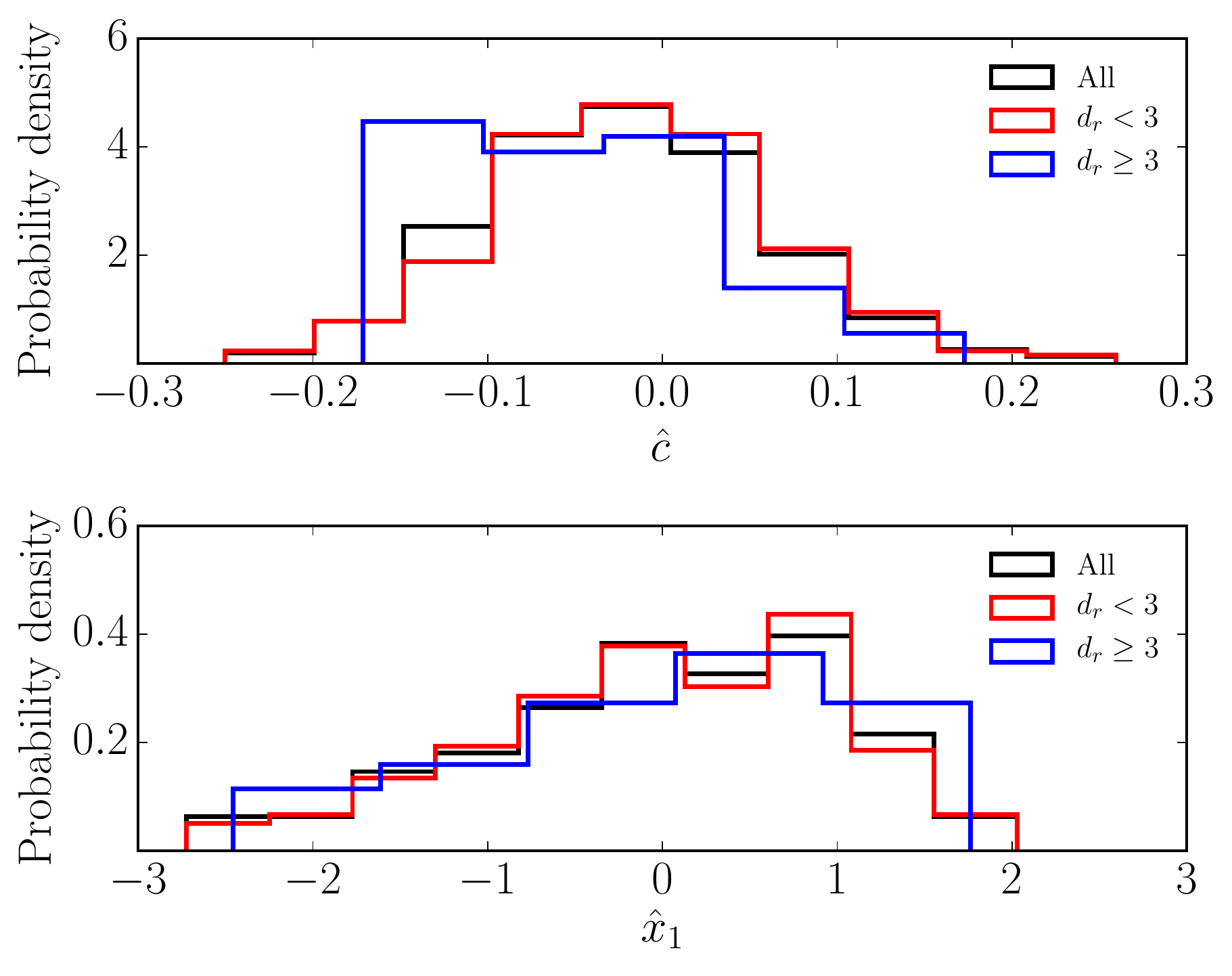}
\caption{Histograms for the observed colour (top) and stretch (bottom) correction. Red is the $\dr < 3$ sub-group, blue is for $\dr \geq 3$, black is for all \SNIas\ combined. The distribution for colour with $\dr \geq 3$ is bluer (i.e., $c < 0$). The distributions for stretch are similar for both sub-groups.}
\label{fig:histograms}
\end{figure}

Using the fixed choice of $\dc = 3$, we define two groups of \SNIas, (i) those with $\dr < 3$ (250 \SNIas) and (ii) those with $\dr \geq 3$ (52 \SNIas). 
The distributions of stretch and colour for the two groups are shown in Fig.~\ref{fig:histograms}. 
The stretch distributions for the two groups are similar, while the $\dr \geq 3$ group appears bluer, i.e., there are more objects with $c<0$.

We use a two-sample Kolmogorov-Smirnov (KS) test to assess the statistical significance of the difference in the colour and stretch distributions of the two groups. The KS test statistic is the maximum difference between the cumulative distribution function of two empirical distributions, $D_\text{max}$. Here we compare the cumulative distributions of the colour for the two groups of \SNIas. $D_\text{max}$ is compared to a threshold value 
\begin{equation}
	D_{n_1, n_2}(\alpha) = \kappa(\alpha)\sqrt{\frac{n_1 + n_2}{n_1 n_2}}
\end{equation}
where $\kappa(\alpha$) is an $\alpha$-dependent critical value and $n_1$ and $n_2$ are the numbers of objects in each of the two groups. The test is significant (at the level $\alpha$) if $\Delta \equiv D_\text{max} - D_{n_1, n_2}(\alpha) > 0$. For $\alpha = 0.05$, the critical level is $\kappa(0.05) = 1.36$.

\begin{figure}
\begin{center}
\includegraphics[width=\linewidth]{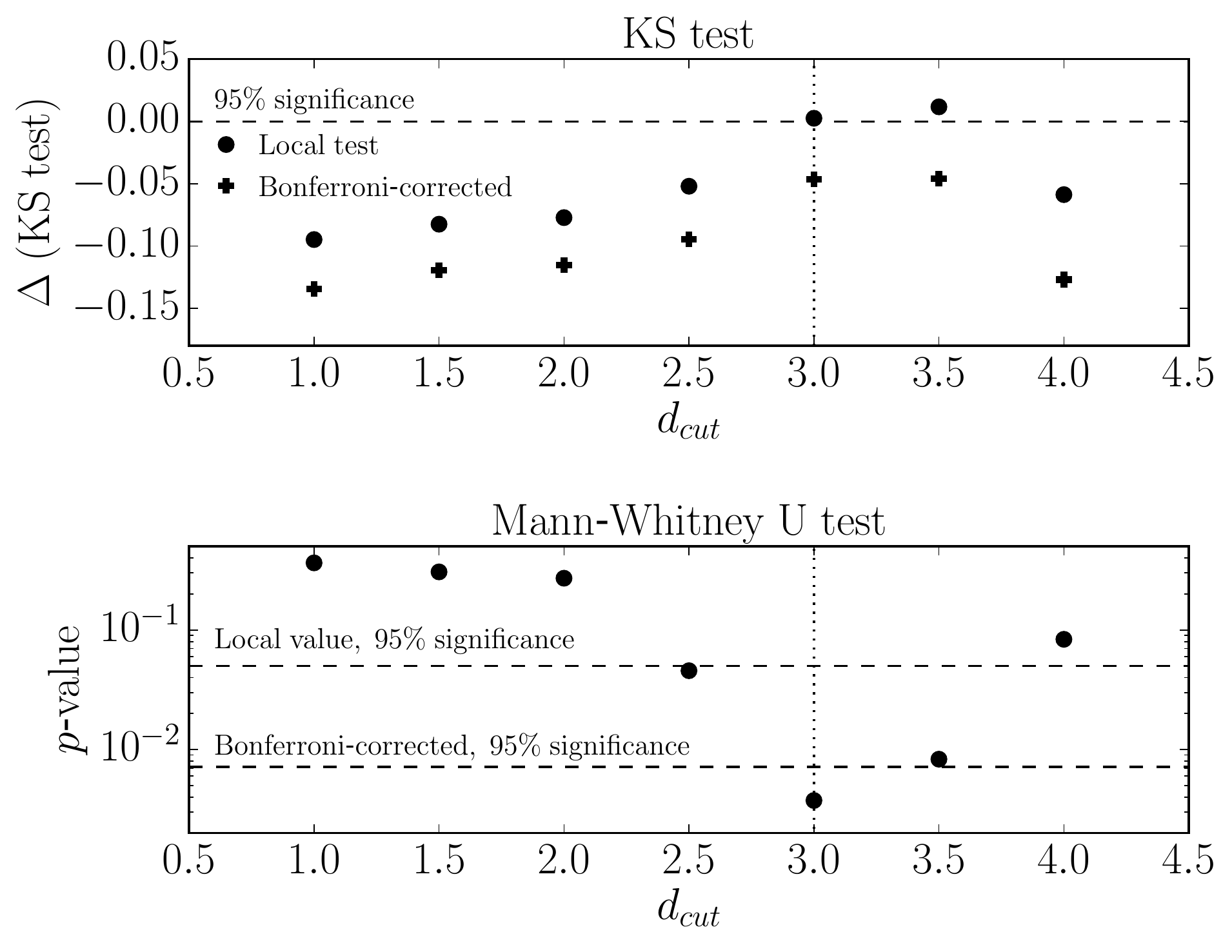}
\caption{Top: Difference between the KS test statistic for the colour correction of two SNIa groups and the significance threshold as a function of the chosen cut-off value for the two \SNIas\ groups, where a value $>0$ (shown the dashed line) indicates a statistically significant difference. Our choice, $\dc=3$, is indicated by the vertical dotted line. The black points were calculated using $\alpha=0.05$ for a significance level of 95 per cent, and the black crosses were Bonferroni-corrected using $\alpha=0.05/7$. Bottom: $p$-values obtained from the two-sample one-sided Mann-Whitney U test as a function of $\dc$, with $\dc=3$ shown as the vertical dotted line. The black points are the calculated $p$-values, the upper dashed line represents a $p$-value of 0.05, and the lower dashed line represents a Bonferroni-corrected $p$-value of 0.05/7.}
\label{fig:KStest}
\end{center}
\end{figure}

To investigate the sensitivity of this result to the choice of $\dc$, we evaluate $\Delta$ as a function of $\dc$ for comparing the distributions of both the colour and stretch corrections in the two SNIa groups. With a significance level of $\alpha=0.05$, the KS test for the stretch distribution is not significant for any value $1 \leq \dc \leq 4$. (We do not consider larger values for $\dc$ as the number of \SNIas\ in the $\dr \geq \dc$ group then becomes too small.)  The KS test is, however, locally statistically significant for the colour distribution for $3 \leq d_{\rm cut} \leq 3.5$, as shown by the black points in the upper panel of Fig.~\ref{fig:KStest}. To account for multiple testing, we use the (conservative) Bonferroni correction, where in order to achieve a confidence level of 95 per cent one must divide $\alpha$ by the number of tests performed. Here we have a total of 7 tests (given by the chosen values of $\dc$), thus we calculate $\Delta$ with a significance level of $\alpha=0.05/7$. The results are shown as crosses in the upper panel of Fig.~\ref{fig:KStest}, and indicate that the two distributions are no longer significantly different at the 95 per cent confidence level when using the Bonferroni correction.

We investigate the choice of $\dc$ further using the non-parametric two-sample one-sided Mann-Whitney U test (sometimes also called the Wilcoxon rank sum test). The hypothesis we want to test is that the sub-group at greater galactocentric distances is bluer than the sub-group at smaller distances. This means that the observed colour distribution for the $\dr \geq \dc$ group should be shifted to the left with respect to the distribution for the $\dr < \dc$ group. This translates into a one-sided hypothesis test. Let $A$ denote a SNIa drawn at random from the $\dr<\dc$ group, and $B$ denote a SNIa drawn at random from the $\dr \geq \dc$ group. We use the Mann-Whitney U statistic to test the one sided null hypothesis that $B$ has a smaller observed colour correction value, $\hat c$, than $A$ with probability greater than or equal to 0.5 against the alternative hypothesis that this probability is less than 0.5. If the null hypothesis is rejected (at a given confidence level), this means that the data indicate that the distribution of the colour correction in the $\dr \geq \dc$ group is to the left of (smaller than) this distribution of
the  $\dr<\dc$ group. This test does not require the colour correction to be normally distributed in the two groups. The one-sided Mann-Whitney U test is generally more powerful than the KS test, and it can detect differences in the shapes of the distributions, not just in their medians (as sometimes incorrectly stated). We show the resulting $p$-values for testing the null hypothesis as a function of $\dc$ in the lower panel of Fig.~\ref{fig:KStest}. The top dashed line corresponds to a $p$-value of 0.05, while the bottom dashed line corresponds to a Bonferroni-corrected $p$-value of 0.05/7. P-values below these lines indicate that the null hypotheses can be rejected at the (local or global) 95 per cent confidence level. We observe that at $\dc=3$ the Mann-Whitney test indicates that the \SNIas\ in the $\dr \geq \dc$ group have a smaller colour correction distribution than do those in the $\dr<\dc$ group. This holds at the 95 per cent confidence level even when the Bonferroni correction is taken into account, in contrast to the KS test.

\section{Evaluating the residual scatter}
\label{sec:BAHAMAS}

After the empirical colour and stretch corrections, \SNIas\ still exhibit a residual scatter in their (standardized) intrinsic magnitudes. Reducing this scatter enables more precise estimation of the cosmological parameters, since it allows for a more precise measurement of the distance modulus as a function of redshift. Here we turn to the question of whether there is a statistically significant difference between the magnitude of the residual scatter in the two subgroups defined in Section~\ref{sec:two_popn}.
We use the Bayesian hierarchical framework \BAHAMAS\ \citep{2011MNRAS.418.2308M,Shariff2016} to estimate the residual scatter of the two groups. This fully Bayesian approach is different from the classical $\chi^2$ approach in two fundamental ways.

Firstly, for each observed variable a latent (true) unobserved value is introduced using a hierarchical representation of the probabilistic relationships between the latent variables, observations and population parameters. More specifically, 
\begin{equation} 
\label{eq:covariates_relation}
\mB{i} =\mu_{i}(\zhat_i, \Cparams)  - \alpha  x_{1i} + \beta  c_i +M_i^{\epsilon}.
\end{equation}
encapsulates the linear corrections to the B-band peak apparent magnitude, $\mB{i}$, captured by the SALT2 correction, namely $c_i$ (colour) and $x_{1i}$ (stretch). Here, $\mu_{i}(\zhat_i, \Cparams)$ is the distance modulus, which depends on the SN's redshift, $\zhat_i $, and the underlying cosmological parameters, $\Cparams$; $\left \{\alpha, \beta \right \}$ are regression coefficients to be estimated, and $M_i^{\epsilon}$ is the residual absolute magnitude after empirical corrections. More specifically, the SALT2 colour parameter is defined as $c = (B-V)_\text{max} - \langle (B-V)_\text{max} \rangle$, where all colours are evaluated at the time of B-band maximum and $\langle (B-V)_\text{max} \rangle$ is the colour averaged over the \SNIas\ in the templates \citep{Guy:2007dv}. A value of $c<0$ means that the blue band magnitude is smaller (i.e., brighter) than the $V$ band, hence the object is bluer. On the other hand, $c>0$ means the object is redder (than average). 
Unlike in the $\chi^2$ approach, in \BAHAMAS\ $\left \{   \mB{i} ,x_{1i} ,c_i  \right \}$ are latent variables that are stochastically related to their observed counterparts, but differ from the data because of observational noise. 
Since the \SNIas\ in our data are spectroscopically confirmed, the errors in redshift can be ignored. The redshift range of the \SNIas\ used in this analysis is restricted by the need to acquire host galaxy images, and spans $0.04<z<0.40$. With this narrow range, SNIa data alone cannot constrain the cosmological parameters.
We therefore assume a flat $\Lambda$CDM universe with a fixed cosmology, $\left \{   \Omega_m = 0.315, H_0 = 67.3 \right \}$, matching the posterior mean value obtained by the {\em Planck} Cosmic Microwave Background mission~\citep{2015arXiv150201589P}.

The second key feature of the Bayesian approach is its ability to account for population variability. Instead of assuming that each SNIa has the same absolute magnitude (after corrections), we assume that they vary probabilistically with an underlying population-level distribution. Specifically, we model the residual absolute magnitude, $M_i^\epsilon$, using a Gaussian distribution:
\begin{equation}
M_i^\epsilon  \sim \normal (M_0, \sigmares^2),
\end{equation}
where $M_0$ is the mean of the residual absolute magnitudes and $\sigmares$ is its residual standard deviation, quantifying the residual scatter of \SNIas\ 
after corrections.  This quantity is to be understood as a phenomenological error describing the intrinsic residual variability that is not accounted for by the empirical corrections for stretch and colour.

It was demonstrated by~\cite{2011MNRAS.418.2308M} with simulated data that (a precursor of) the Bayesian approach used in \BAHAMAS\ results in less biased, more accurate estimates than the standard $\chi^2$ approach, while improving overall coverage of the ensuing credible intervals.  We apply \BAHAMAS\ to the two SNIa groups ($\dr <3$ and $\dr\geq3$), as well as to the whole dataset for comparison, and obtain the marginal posterior distribution of $\left \{\alpha, \beta, \sigmares \right \}$ numerically using the sampler presented in \cite{Shariff2016}. Full details of \BAHAMAS, including prior choices, hierarchical structure and the sampling algorithms are given in~\cite{Shariff2016}.

Our approach to estimating the residual scatter, and the colour and stretch corrections parameters is significantly more sophisticated than what has been adopted in previous works attempting to establish the influence of galactocentric distance~\citep{ivanov2000,galbany2012}. The Bayesian method adopted here has the advantage of exploiting the probabilistic nature of the model to `borrow strength' between \SNIas, thus increasing sensitivity to subtler features in the data than relative to the cruder $\chi^2$ approach. Furthermore, we produce a full marginal probability distribution for the value of the intrinsic scatter, not just a point estimate. As shown in~\cite{2011MNRAS.418.2308M} this results in more accurate and precise estimates for the parameters of interest than it is possible with the standard $\chi^2$ method.  

In the present work we segregate \SNIas\ according to the two sub-groups defined above. 
Alternatively, one could include $\dr$ as an additional linear covariate in the correction term in Eq.~\eqref{eq:covariates_relation}, the slope of which would be another free parameter, thus replacing  Eq.~\eqref{eq:covariates_relation} with:
\begin{equation} 
\label{eq:covariates_relation_general}
\mB{i} =\mu_{i}(\zhat_i, \Cparams) + X_i^T \mathcal{B}  +M_i^{\epsilon},
\end{equation}
where $X_i = \{x_{1i}, c_i, d_{r,i}\}^T$ and $\mathcal{B} = \{-\alpha, \beta, \gamma\}^T$, with $\gamma$ the slope of the galactocentric distance covariate. One would then fit the value of $\gamma$ together with the colour and stretch correction coefficients, $\alpha$ and $\beta$.  The inclusion of galactocentric distance as an additional standardization variable is akin to how the host galaxy mass correction is usually parameterized. In both cases, one can either segregate the \SNIas\ in two groups, divided by a cut-off value, or use the additional covariate as a linear correction term in the distance modulus, as was done e.g., in~\cite{Shariff2016}. The second approach dispenses with the need of defining an arbitrary cut-off point. 

In previous studies,~\cite{ivanov2000,galbany2012} investigated the impact of galactocentric distance as a covariate. \cite{galbany2012} also used only two bins (`near' and `far'). Importantly, the `near' and `far' bins were defined by splitting the \SNIas\ into two groups of  equal size (and further subdividing them according to host morphology), and not with reference to their normalized galactocentric distance, like we do here. This might have washed out any potential correlation. A puzzling result of \cite{galbany2012} is that while $c$ decreased with distance (with a reported significance of the correlation coefficient of 4.4$\sigma$) for the entire data set, the effect disappeared when the data were split according to morphology. 
\cite{Yasuda:2009sc} also adopted (de-projected) galactocentric distance as a covariate, finding no effect. They, however, did not attempt to normalize the galactocentric distance to the host light radius, as we do here. 

For simplicity, we only consider splitting the \SNIas\ into two groups (with a hard, pre-defined cut), and leave investigation of the galactocentric distance as a linear covariate to future work.

\section{Results}
\label{sec:results}

\begin{figure}
\centering
	\includegraphics[width=\linewidth]{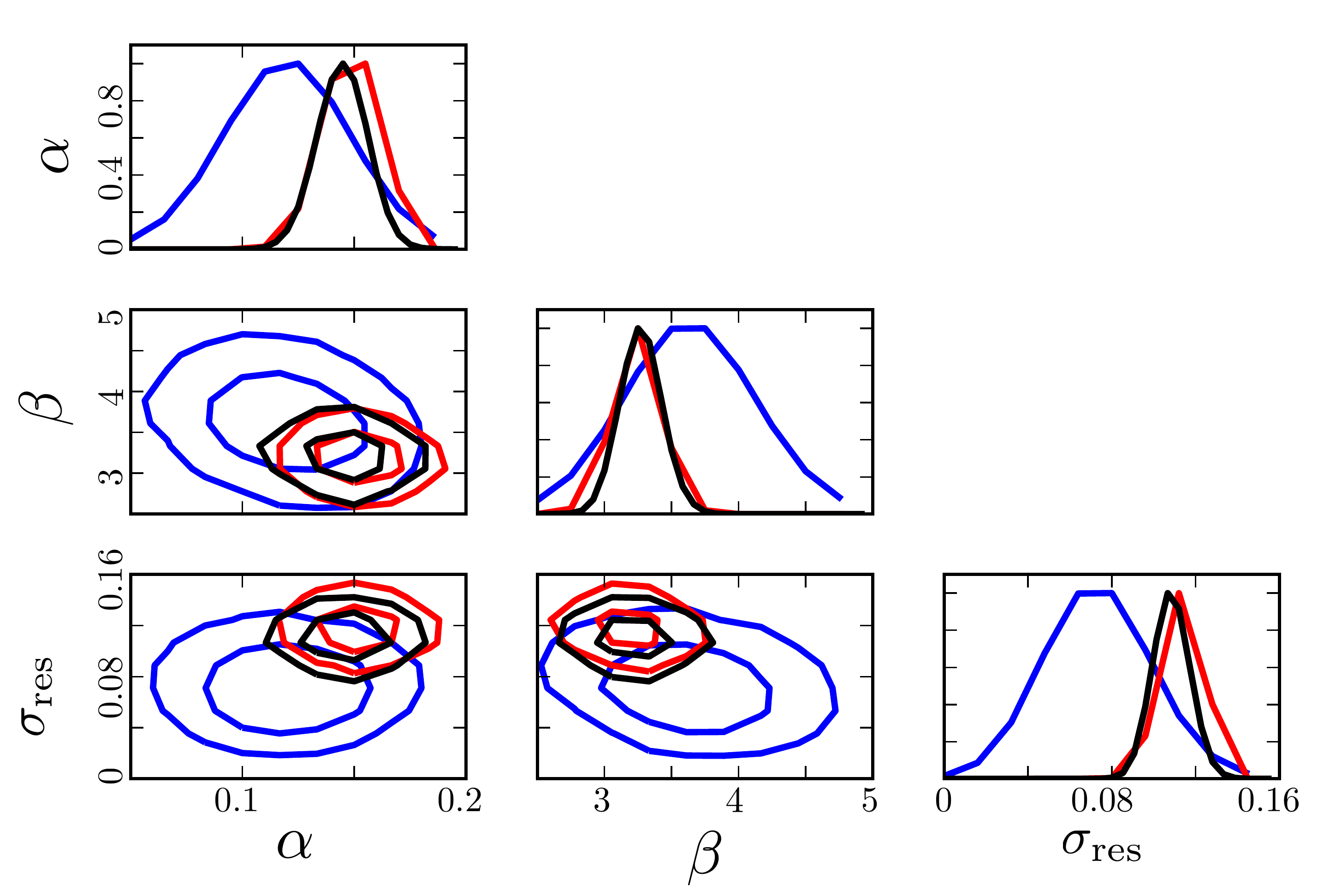}
\caption{1D and 2D marginal posterior distributions for $\alpha$, $\beta$ and $\sigmares$ (the residual scatter in the intrinsic magnitude), with all latent variables and other population-level parameters marginalized over. Red is for the sub-group at small galactocentric distance, $\dr<3$ ($N_\text{SNIa} = 231$), blue is for the group at high galactocentric distance, $\dr\geq 3$ ($N_\text{SNIa} = 49$), while black is all \SNIas\ combined. 1D posterior densities have been normalized to the peak. 2D contours depict 68 per cent and 95 per cent credible regions.}
\label{fig:posterior}
\end{figure}

The marginal posterior distributions in one and two dimensions for the parameters of interest, $\left \{\alpha, \beta, \sigmares \right \}$ are shown in Fig.~\ref{fig:posterior}, split according to sub-group and compared with the result for the whole data set. Table \ref{table:posterior} gives posterior summary statistics for the same quantities. In both cases, the latent parameters $\{ c_i, x_{1,i}, \mB{i}\}$ ($i=1, \dots , N$, where $N=302$ is the number of \SNIas\ considered) and all other population-level parameters, $\{ c_*, R_c, x_*, R_x \}$, have been marginalized out. Here, $c_*$ and $x_*$ are the (redshift-independent) population means of the colour and stretch distributions, respectively, and $R_c, R_x$ are their variances. The distributions of colour and stretch are taken to be Gaussian, see ~\cite{Shariff2016} for full details. 

\begin{table*}\footnotesize
\begin{center}
\begin{tabular}{lccc} 
\hline\hline 
& $\dr<3$ & $\dr\geq3$ & All data \\
\hline 
$\alpha$ &$0.149 \pm 0.012$ & $0.120 \pm 0.020$ & $0.145 \pm 0.010$    \\ 
$\beta$ & $3.244  \pm 0.162$ & $3.640  \pm 0.353$ & $3.271  \pm 0.142$   \\ 
$\Mnote$ & $-19.116\pm 0.065$ & $-19.152 \pm 0.049$ & $-19.145 \pm 0.046$ \\ 
 $\sigmares$  & $0.114 \pm 0.009$ & $0.073 \pm 0.018$ & $0.108 \pm 0.008$\\ 
\hline \hline
\end{tabular}
\caption{Marginal posterior average and standard deviation for the stretch, $\alpha$ and colour, $\beta$, correction parameters, as well as for the \SNIas\ residual scatter $\sigmares$ (after empirical corrections) for the three cases considered: including only \SNIas\ with small galactocentric distance ($\dr < 3$), only \SNIas\ with large galactocentric distance ($\dr \geq 3$) and including all \SNIas\, irrespective of galactocentric distance (`All data').}
\label{table:posterior}
\end{center}
\end{table*}

\begin{figure}
\begin{center}
\includegraphics[width=\linewidth]{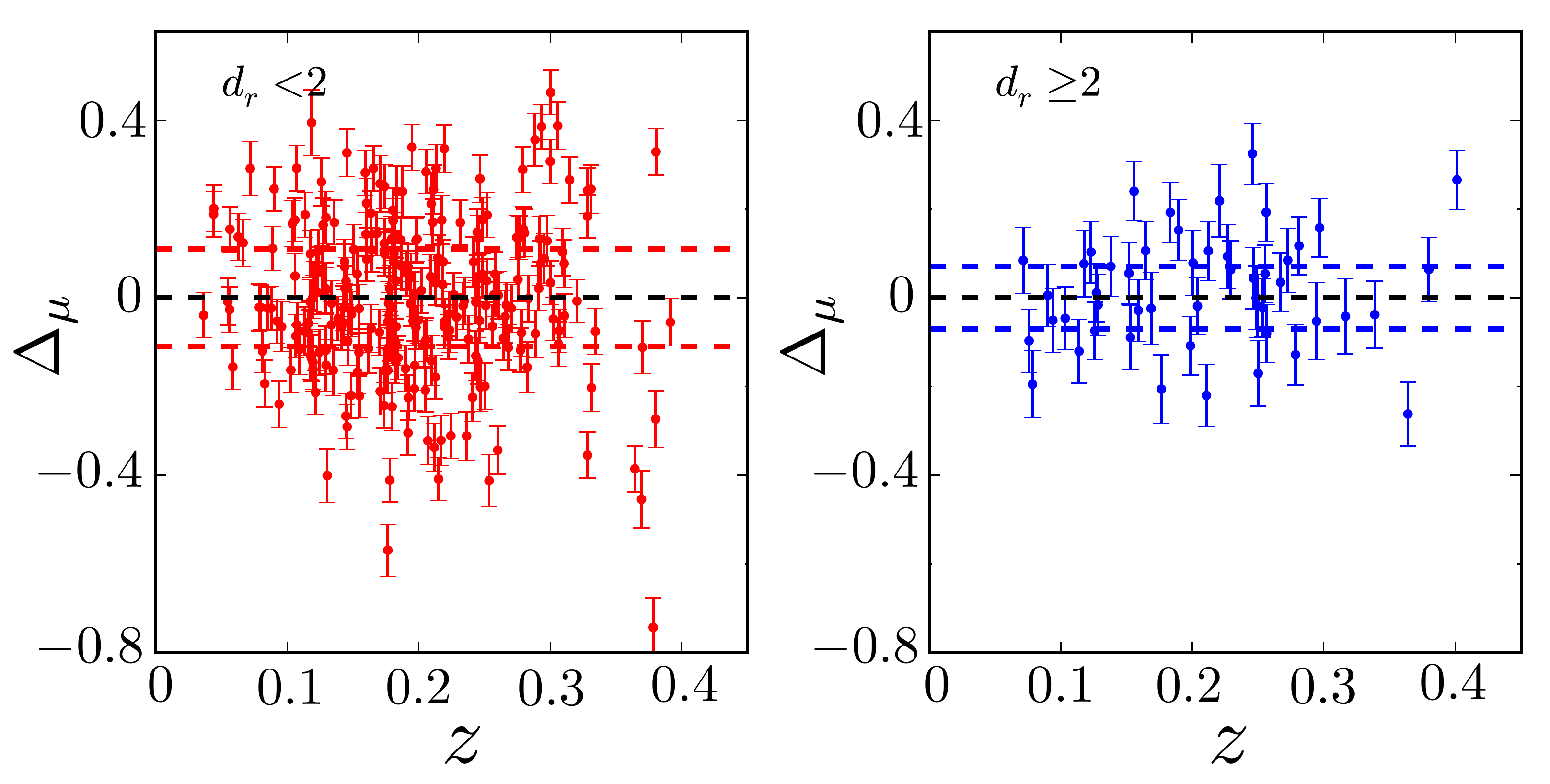}
\caption{Hubble residuals as a function of redshift. The left panel shows the residuals for the \SNIas\ with $\dr<3$, while the right panel shows \SNIas\ with $\dr\geq 3$. Error bars are the posterior standard deviation of the Hubble residual value. Also shown as horizontal dashed blue/red lines is the posterior mean of the residual scatter, $\sigmares$, given in Table~\ref{table:posterior}.}
\label{fig:hubble_residual}
\end{center}
\end{figure}

We observe shifts in the distributions of the all quantities for the two different sub-groups. The most dramatic difference is in the value of the residual scatter, $\sigmares$. Its posterior distribution is fairly Gaussian, and has posterior average and standard deviation  given by $\sigmares = 0.114 \pm 0.009$ for the $\dr < 3$ sub-group, which is reduced to  $\sigmares =0.073 \pm 0.018$ for the $\dr \geq 3$  sub-group. This represents a reduction of $\approx 30$ per cent in the posterior mean values. The significance of the difference is approximately 2.0$\sigma$ (computed using the standard deviation of the difference between $\sigmares$ in the two subgroups and assuming Gaussian errors). While this is not strongly significant, we emphasize the small sample size ($N_\text{SNIa} = 49$ for the $\dr \geq 3$ sub-group), which results in a fairly wide posterior distribution for $\sigmares$. This means that \SNIas\ at large galactocentric distances show better standardization properties (i.e., are more uniform in their magnitudes after corrections) than the  whole data set.
This is additionally supported by Fig.~\ref{fig:hubble_residual}, showing the Hubble residuals of the two sub-groups. The right/blue ($\dr\geq3$) residuals are clearly smaller than the left/red ones ($\dr<3$). \SNIas\ further away from the centre of the galaxy have, on average, smaller residuals after corrections. Therefore, cosmological distance estimation from this sub-group can be expected to be more precise. 

\begin{figure}
\centering
\includegraphics[width=\linewidth]{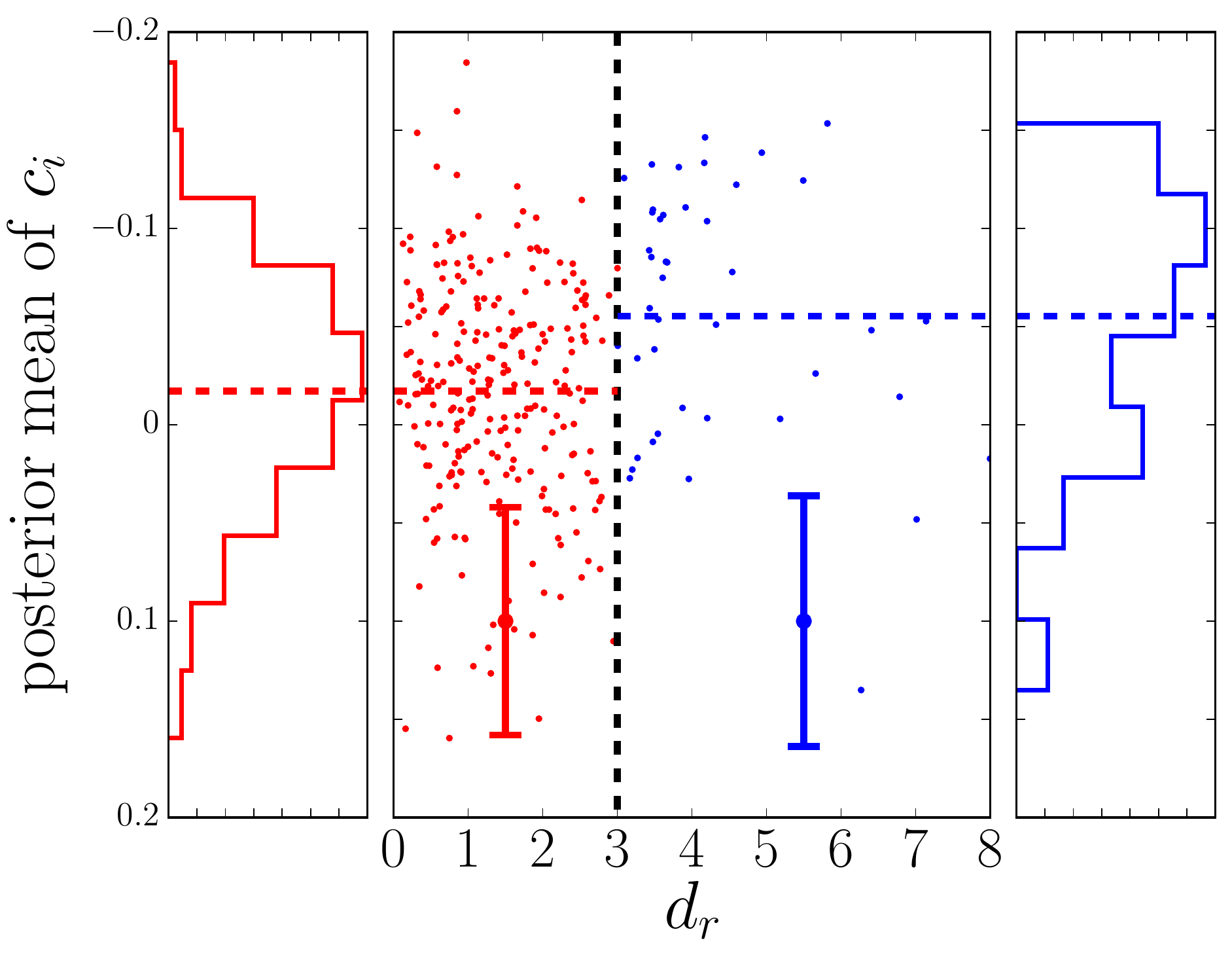}
\caption{Posterior mean of the latent colour values (determined from our hierarchical model) as a function of galactocentric distance, $\dr$, colour coded according to the two sub-groups (split at $\dr=3$). The histograms at the margins show the 1D marginal distribution of latent colour values for the two subgroups. The $\dr \geq 3$ sub-group clearly shows a preferentially bluer colour ($c<0$). The vertical errorbars give the average posterior standard deviation in the values of $c_i$ for each of the two sub-groups, and the dashed horizontal lines indicate their means.
\label{fig:latent_colour}}
\end{figure}

\begin{figure}
\centering
	\includegraphics[width=\linewidth]{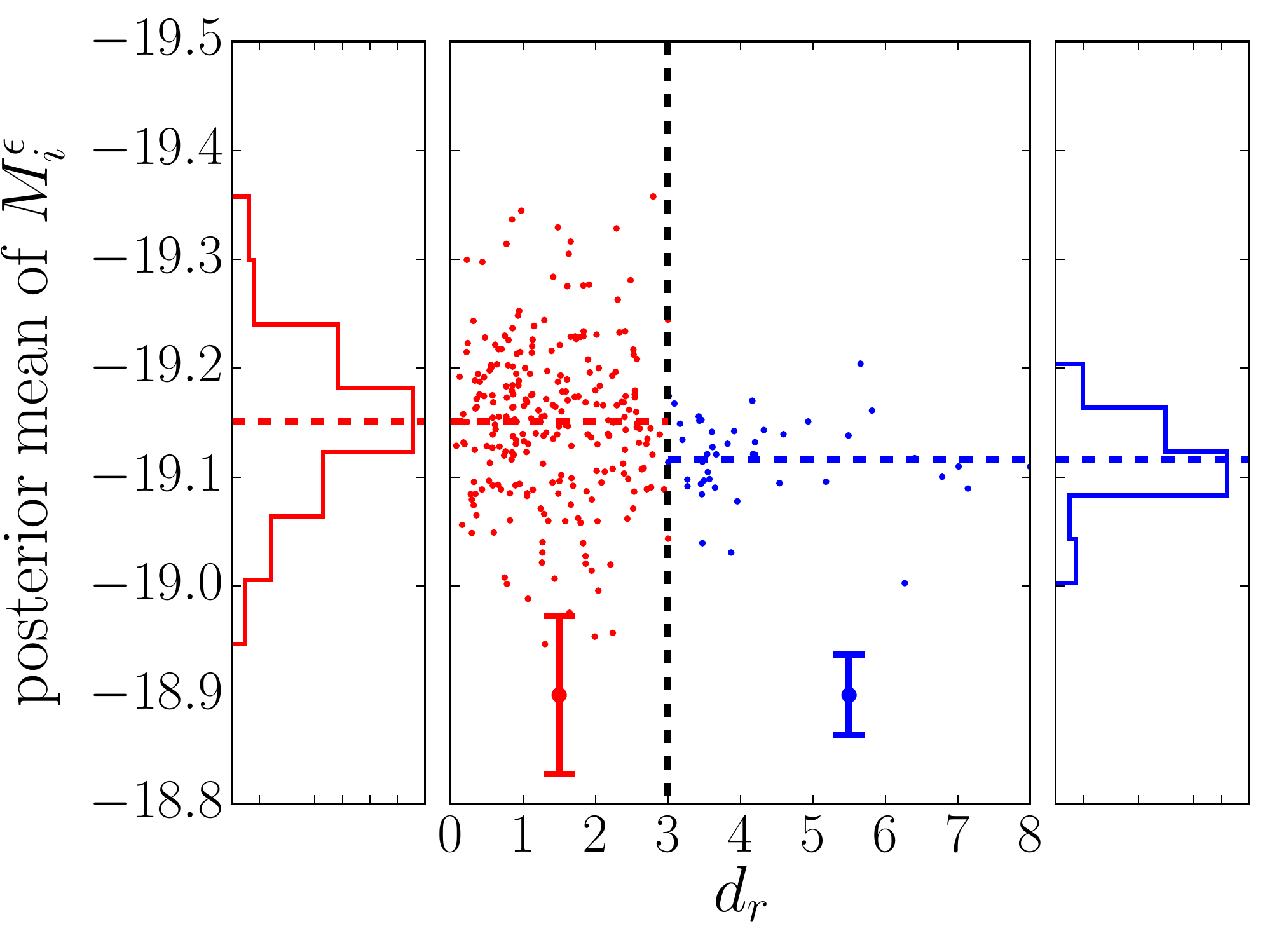}
\caption{Posterior mean of the (post-correction) intrinsic magnitude for the two sub-groups. \SNIas\ at large galactocentric distance ($\dr \geq 3$) exhibit a tighter distribution around the mean magnitude. The vertical errorbars give the average posterior standard deviation in the values of $M_i^\epsilon$ for each of the two sub-groups, and the dashed horizontal lines indicate their means.
\label{fig:latent_M}}
\end{figure}

We show in Figures~\ref{fig:latent_colour} and~\ref{fig:latent_M} the inferred posterior mean and standard deviation for the latent colour and (post-correction) intrinsic magnitude for each SNIa, fit separately for the two subgroups. Fig.~\ref{fig:latent_colour} shows that indeed the $\dr \geq 3$ sub-group has a latent colour distribution that is skewed towards bluer objects (we note that this result still holds even for the observed colours, albeit with a larger scatter). From Fig.~\ref{fig:latent_M} we observe that the  $\dr \geq 3$ sub-group distribution of intrinsic magnitudes is tighter. {The difference in mean absolute magnitude (post-correction) between the two groups is $-0.035 \pm 0.082$, so compatible with zero. Therefore, we conclude that there is no evidence for a difference in the average, post-correction magnitudes in the \SNIas\ between the two sub-groups.}

One could imagine that the physical explanation for our findings might lie in the reduced amount of dust found at high galactocentric radii: since the observed colour variation is the sum of intrinsic colour variability and reddening due to the host galaxy and/or local circumstellar material (CSM) dust (which varies from host to host), reducing the host galaxy colour variability should lead to a smaller residual scatter after correction, as observed in our results. {However, this interpretation is at odds with the inferred value of the colour correction coefficient for the $\dr \geq 3$ sub-group ($\beta = 3.640  \pm 0.353$), which is larger than for the $\dr < 3$ sub-group ($\beta = 3.244  \pm 0.162$). Owing to the larger uncertainty in the $\beta$ value for the $\dr \geq 3$ sub-group (which contains a smaller number of \SNIas), the difference is not statistically significant. However, the fact that the $\dr \geq 3$ sub-group has a larger posterior mean for $\beta$ cannot be reconciled with the notion that high galactocentric distance \SNIas\ are subject to reduced host galaxy reddening, for the following reason.} 

There is evidence \citep{Wang:2005tc,Goobar:2008qp} that the local CSM might have smaller size dust grains (compared to the Milky Way), leading to a smaller value of the total-to-selective extinction $R_V \equiv A_V/E(B-V)$~ than the average Milky Way value of $R^\text{MW}_V = 3.1$. Here, $A_X$ is the extinction (in mag) in band $X$, i.e., $A_X = m_X - m_{X,0}$, where $m_X$ is the apparent magnitude in band $X$ and $m_{X,0}$ is the apparent magnitude in the same band in the absence of extinction. The colour excess $E(B-V)$ is given by $E(B-V) = (m_B-m_V) - (m_{B,0}-m_{V,0}) = A_B - A_V$. Given that total extinction and reddening is the cumulative effect of local CSM dust and ISM dust, if the effect of the ISM were largely removed by the cut to large $\dr$, one would expect this sub-group to exhibit the reddening law of the underlying CSM, which is typically steeper\footnote{The wavelength dependency of extinction in the vicinity of the $V$ band is approximately linear in $1/\lambda$ with slope $1/R_V$, where $\lambda$ is the wavelength~\citep{Cardelli:1989fp}. Hence a smaller value of $R_V$ results in a steeper reddening law.} than the Milky Way value. For example, for SN2014J~\cite{Yang:2016qna} found evidence of a luminous arc -- attributed to the CSM -- with an estimated $R^\text{CSM}_V \approx 1.4$.  Recall that, from Eq.~\eqref{eq:covariates_relation}, $\beta$ is the slope that gives the change in $B$ band magnitude for a unit change in the value of $B-V$. Therefore, $\beta$ ought to be compared to $R_B \equiv A_B/E(B-V)$. Given their respective definitions, it follows that $R_B-R_V = (A_B-A_V)/E(B-V)= 1$, hence $R_B = R_V + 1$. From this, one would expect that reddening due to the CSM should typically show $\beta^\text{CSM} = R_V^\text{CSM} + 1 \approx 2.4$, i.e., a much smaller value than we observe in the $\dr \geq 3$ sub-sample, {for which $\beta = 3.64 \pm 0.35$}. 

Alternatively, in the absence of CSM reddening one would expect the inferred value of $\beta$ for the $\dr \geq 3$ sub-sample to recover the value associated with the intrinsic colour variability of the \SNIas. This has been estimated by \cite{Mandel:2016rks} in a low-redshift sample, finding $\beta^\text{int} = 2.210\pm 0.255$, again much lower than the value we obtain for our sub-sample at high galactocentric distance. \cite{Mandel:2016rks} argue that the simple linear colour correction formula (which we adopt in this work, in line with standard usage) effectively averages between the intrinsic colour correction, $\beta^\text{int}$, and the reddening law slope. Thus one would expect that if indeed the $\dr \geq 3$ sub-group had negligible reddening due to the ISM, we would observe a value of $\beta$ that averages between the $\beta^\text{int} = 2.210$ value found by ~\cite{Mandel:2016rks} and the CSM reddening law, equivalent to $\beta \approx 2.2-2.4$. Thus the inferred $\beta$ for the $\dr \geq 3$ sub-group ought to be around 2.2, which is not what we observe.

\cite{2010AJ....139..120F} analysed the first set of low-redshift \SNIas\ from the Carnegie Supernovae Project (CSP), and found $R_V \approx 1.7$ ($\beta \approx 2.7)$ for the entire set of \SNIas. However, when excluding two highly reddened \SNIas\ from their analysis, they obtained $R_V \approx 3.2$ ($\beta \approx 4.2)$. While their value is somewhat higher than our value for the $\dr \geq 3$ sub-group ({$\beta = 3.640 \pm 0.353$}), the effect goes in the same direction. Our value for $\beta$ for the larger galactocentric distance sub-group however still falls short of the average Milky Way value of $R_V = 3.1$ (corresponding to $\beta = 4.1$).
More recently, \cite{2014ApJ...789...32B} used the late-time \cite{1996MsT..........3L} law to select a low-reddening sample of 34 \SNIas\ from the Carnegie Supernovae Project (CSP). Their approach circumvents the use of galactocentric distance, which can be expected to assign to the $\dr < 3$ sub-group \SNIas\ that are low reddened but at small projected radii (i.e., in front of the host). \cite{2014ApJ...789...32B} then employed a Bayesian hierarchical model to reconstruct the reddening law of the host (under a number of priors). They found that objects with the least reddening ($E(B-V)_\text{host} < 0.2$) have a reddening law compatible with the Milky Way value of $R_V = 3.1$ (corresponding to $\beta=4.1$). This is compatible with our observation of a shift of $\beta$ to larger values in the $\dr\geq 3$ group, which (being the bluer sub-group) exhibits the least reddening. This thus hints at the possibility that the shift of $\beta$ to higher values for the $\dr \geq 3$ sub-group is compatible with the hypothesis that the remaining reddening is due to the host ISM.  Furthermore, despite not finding any correlation between colour and galactocentric distance, \cite{Yasuda:2009sc} do observe an excess of redder colour values at low galactocentric distances,  compatible with the later findings of \cite{Anderson:2014aaa}. Another heuristic (if counterintuitive) argument in support of our findings comes from the fact that less massive galaxies (with stellar mass $M_* < 10^{10} M_\odot$) show a larger value of $\beta$ \citep{Sullivan:2010mg}. These galaxies are also the ones with the least dusty environments~\citep{Garn:2010ws}. Hence a larger value of $\beta$ can be associated with less dusty environments.

Taken all together, the above arguments lend credence to the hypothesis that the large $\dr$ sub-group is actually probing a low-reddening/low-attenuation ISM environment, rather than seeing the underlying CSM/intrinsic colour scatter. The reduction in residual intrinsic scatter we observe could be the resultant of the $\dr \geq 3$ sub-group being subject to a less dusty ISM, as well as to the more homogeneous metallicity and SFR environment for this sub-group. Further work is required to disentangle the physical origin of this effect, and we comment on this aspect in the next section.

In this work we have not divided hosts in terms of their morphology. Since early-type galaxies are dominated by old stars and relatively dust-free~\citep{1996ApJ...461..155W}, while spirals have star-forming activity in the arms and a diverse (age-wise) stellar population, \SNIas\ that occur in early-type galaxies are less likely to be significantly affected by dust. In spirals, central stars tend to be both older and metal rich than in outer regions, and therefore classifying \SNIas\ according to both radial distance and galaxy type could potentially disentangle the effects due to stellar age and metallicity from those due to dust properties. One might therefore speculate that including host galaxy morphology information could help in further reducing the residual scatter, and perhaps in clarifying the origins of the effect. \cite{lampeitl2010} subdivided the SDSS-II SNIa sample in two groups, according to the passive or star forming nature of their host galaxies (based on their estimated SFR). They found a strong difference in the stretch parameter, $x_1$, for the two groups, but no significant difference in the colour, concluding that \SNIas\ must have the same intrinsic colour variations in all galaxy types. This conclusion was confirmed by \cite{Henne:2016mkt}, who examined the influence of galaxy morphology by classifying hosts in three groups -- ellipticals/lenticulars (E,S0), early-type spirals (Sa-Sc), late-type spirals (Sd-Ir)-- for 192 \SNIas\ from JLA. While they reported 1$\sigma$ shifts in the reconstructed values of $\alpha, \beta$ depending on galactic morphology group, they did not find any significant correlation between colour and host morphology. This conclusion stood despite a weak trend of slightly bluer colour \SNIas\ in early type (i.e., passive) hosts. This was interpreted as being due to the larger amount of dust in spirals, which would thus make the SN colour redder. The above studies thus seem to suggest that our results are not strongly affected by the lack of morphological host information.

 {Even the spectroscopic SDSS-II SN sample is known to be complete only out to $z\sim 0.2$ \citep{sako2018}, and we have used \SNIas\ out to $z=0.4$. As the apparent magnitude of high-$z$ \SNIas\ approaches the flux limit of the telescope, brighter objects will be detected preferentially, leading to a bias in their inferred distance modulus. The $d_r \geq 3$ sub-group has on average smaller colour values, and an inferred larger $\beta$ value than the $d_r < 3$ group, which means that the colour correction term $\beta c_i$ in Eq.~\eqref{eq:covariates_relation} for the $d_r \geq 3$ sub-group is (at a given redshift) more negative than for the $d_r < 3$ sub-group. It follows from Eq.~\eqref{eq:covariates_relation} that at a fixed redshift (and thus fixed value of $\mu_i(\hat{z_i}, \mathcal{C})$) the apparent magnitude for the $d_r \geq 3$ sub-group is on average smaller than for the $d_r < 3$ sub-group (since there are no significant differences in either $\alpha x_{1i}$ or $M_i^\epsilon$ between the two groups, as shown in Figures~\ref{fig:histograms} and \ref{fig:latent_M}). Hence for $z>0.2$ \SNIas\  in the $d_r \geq 3$ sub-group can be expected to have a larger apparent brightness and therefore might be preferentially selected due to magnitude-based selection effects. When plotting the ratio of the number of \SNIas\ in the $d_r \geq 3$ sub-group to the $d_r < 3$ sub-group as a function of redshift, we do observed an increase for $z>0.2$. However, due to the smaller number of objects at larger redshift, the Poisson uncertainty in the ratio is quite large, and the increase in the ratio is not statistically significant. Furthermore, when considering the Hubble residuals as a function of $z$ for the two subgroups, we do not observe any feature in their distribution (like a skewness, which would be indicative of residual selection effects) above $z=0.2$. Therefore, we conclude that magnitude-based selection effects do not play a statistically significant role in our results.}

This work does not consider colour-based selection effects over and above the (magnitude) corrections already implemented in the JLA sample. There is evidence that high-$z$ SNIa are bluer~\citep{Rubin:2016iqe} and come from less dusty environment~\citep{Mandel:2016rks} due to selection bias. However, our $\dr \geq 3$ sub-group spans almost the entire SDSS-SNIa redshift range (see Fig.~\ref{fig:distr_dr}), with the majority of the \SNIas\ at $z<0.3$, and hence colour-based selection effects are unlikely to be playing a major role here.

\section{Conclusions}
\label{sec:conclusion}

We have measured the projected galactocentric distance to the host of the \SNIas\ for a sub-set of the SDSS Type Ia \SNIas\ in the JLA sample and investigated the scatter around the Hubble diagram (for a fixed cosmology) for two sub-groups, separated by a cut in their distance. The rationale was that \SNIas\ at large galactocentric distances might be less subject to ISM reddening and absorption, and might explode in more homogeneous environments in terms of metallicity and local SFR. 

We have demonstrated that \SNIas\ further away from the host can be standardized to a higher degree, in that their intrinsic dispersion (post stretch and colour corrections) is reduced from $\sigmares = 0.114 \pm 0.009$ (for the full sample) to $\sigmares = 0.073 \pm 0.018$ (for the \SNIas\ with larger galacocentric distances).  The statistical significance of the effect is, however, small (about $2\sigma$). This is primarily due to the smallness of the sample (only 49 \SNIas\ in the $\dr \geq 3$ sub-group).  Future surveys and data tabulations should be encouraged to include measured galactocentric distances (and, perhaps, morpholgy of host) to account for this effect. In a Bayesian framework, it would be straightforward to include prior information on extinction and reddening depending on galactocentric distance and morphology of host. This has the potential to improve accuracy and precision for inferred cosmological parameters.

Future work should focus on verifying whether a larger sample size can confirm this tentative result. The SDSS-II SN sample contains a larger number of photometrically observed SNIa, which have not been included in this analysis. \cite{2016MNRAS.457.3470C} analysed 721 \SNIas\ from this larger data set, and used different host properties such as age and metallicity to reduce the Hubble residual. Given the availability of SDSS imagery for the hosts, the additional supernovae in the photometric SDSS sample could be analyzed to corroborate or disprove this result. Also, since this is a photometric sample, there is a possibility of contamination of the sample by non-SNIa's. A Bayesian supernova classifier~\citep{2012ApJ...752...79H,Jones:2016cnm,Esben} can be used to account for this. Additionally, since this photometric sample contain SNIa to a higher redshift, selection effects may play a major role and have to be correctly accounted for (e.g. as in \citealt{2015ApJ...813..137R}).  Another, complementary low-redshift ($0.01 < z < 0.10$) \SNIas\ sample is the one used in~\cite{Mandel:2016rks}, from a compilation of data including high-quality light curves from the CfA and CSP surveys. An advantage of this lower redshift data set is that detailed morphological studies can be carried out on these relatively nearby galaxies, to verify whether or not host galaxy type plays a role in the correction procedure. Furthermore, the source of the difference in $\beta$ value from the two sub-groups could be elucidated by analyzing them with the SIMPLE-BayeSN framework~\citep{Mandel:2016rks}. The upcoming SNIa data from the Dark Energy Survey would also be a very useful testing ground for our method. We leave exploration of all these avenues to future work.

{\it Acknowledgements:} we dedicate this paper to the memory of our colleague and co-author, L.B. Lucy (1938-2018). We thank Heather Campbell, Kaisey Mandel and Marc Sullivan for useful discussions. We are grateful to an anonymous referee for helpful suggestions and comments that have improved this work. This work was supported by Grant ST/N000838/1 from the Science and Technology Facilities Council (UK). RT was partially supported by an EPSRC ``Pathways to Impact'' grant.  RT, DvD, and HS were supported by a Marie-Skodowska-Curie RISE (H2020-MSCA-RISE-2015-691164) Grant provided by the European Commission.

\input{DLR_paper.bbl}
\bsp 
\label{lastpage}

\end{document}